\documentclass[hidelinks,onefignum,onetabnum]{siamart220329}

\usepackage{verbatim}
\usepackage{amsmath}
\usepackage{amssymb}
\usepackage{graphicx}
\usepackage{gensymb}
\usepackage{fancybox}
\usepackage{listings}
\usepackage{color}
\usepackage{enumerate}
\usepackage{marginnote}
\usepackage{breqn}
\usepackage{mathrsfs}
\usepackage{cite}

\usepackage{todonotes}

\newcommand{\red}[1]{\ifmmode\mathbf{\textcolor{red}{#1}}\else \textbf{\textcolor{red}{#1}}\fi}

\newcommand{\blue}[1]{\ifmmode{\textcolor{blue}{#1}}\else {\textcolor{blue}{#1}}\fi}

\newcommand{\green}[1]{\ifmmode\mathbf{\textcolor{green}{#1}}\else \textbf{\textcolor{green}{#1}}\fi}

\newcommand{\lbl}[1]{\label{#1}}

\def\K{K}
\def\ba{\beta_0}

\def\eps{\epsilon}

\def\peps{\mathcal{P}_{\max}}

\def\p0{\bar{\mathcal{P}}}

\def\pstar{\mathcal{P}_*}
\def\pmax{\peps}
\def\pmaxh{\hat{\mathcal{P}}_{\max}}

\def\bra{\langle}
\def\ket{\rangle} 
\def\pavg{\bra \pps\ket}

\def\pps{\mathbf{P}}
\def\X{\mathbf{X}}

\def\ppspacing{\lambda}

\def\Dmin{\hat{\mathcal{D}}_{\min}}

\def\Hd{\bar{h}}
\def\Hd{\bar{\mathcal{H}}}

\def\Jnc{J^{nc}}

\def\Jncl{J^{nc}_{\ell}}

\def\non-conservative{\red{non-conserved}}

\def\Hc{\bar{H}_c} 
\def\Hm{\bar{H}_m} 

\def\M{\mathcal{M}}
\def\E{\mathcal{E}}

\def\GLR{\gamma}  
\def\bb{{\mathcal{G}}} 

\def\pinf{P_L} 

\usepackage{hyperref}
\hypersetup{
    colorlinks,
    citecolor=black,
    filecolor=black,
    linkcolor=black,
    urlcolor=black}

\headers{Thin film coarsening with weak condensation}{H. Ji and T. P. Witelski}
\title{Coarsening of thin films with weak condensation}
\author{Hangjie Ji\thanks{Department of Mathematics, North Carolina State
University, Raleigh, NC 27695-8205, USA \email{hangjie\_ji@ncsu.edu}} \and 
Thomas Witelski\thanks{Department of Mathematics, Duke University, Durham,
NC 27708-0320, USA \email{witelski@math.duke.edu}}}

\begin{document}
\maketitle

\begin{abstract}
A lubrication model can be used to describe the dynamics of a weakly volatile viscous fluid layer on a hydrophobic substrate. Thin layers  of the fluid  are unstable to perturbations and break up into slowly evolving interacting droplets. 
A reduced-order dynamical system is derived from the lubrication model based on the nearest-neighbor droplet interactions in the weak condensation limit. Dynamics for periodic arrays of identical drops and pairwise droplet interactions are investigated which provide insights into the coarsening dynamics
for large systems.
Weak condensation 
is shown to be a singular perturbation, fundamentally changing the long-time coarsening dynamics for the droplets and the overall mass of the fluid in two additional regimes of long-time dynamics.
\end{abstract}

\section{Introduction}
The dynamics of viscous fluids spreading on water-repellent surfaces are of interest in connection with many natural and engineering settings \cite{Becker2003}. In what is broadly called dewetting dynamics, nearly uniform thin layers become unstable and break up through several stages of intermediate dynamics to eventually yield large numbers of fluid droplets. 
Lubrication models \cite{oron1997long,craster2009dynamics} have been successfully used to  describe these physical systems using PDE models \cite{bertozzi2001dewetting,witelski2020}.
Locally, each drop will be close to an equilibrium state, but globally the system will exhibit slow dynamics on long timescales 
that
occur due to interactions of the droplets. The edges of adjacent drops can {collide} leading to them to merge together or large drops can grow by drawing fluid from smaller ones, causing the latter to {collapse}. Both of these mechanisms lead to a monotone decrease in the number of drops present as a function of time, $N(t)$; this is often called \emph{coarsening}. Heuristic scaling arguments were used to obtain power-law behaviors for $N$, $N(t)=O(t^{-2/5})$ for the one-dimensional problem \cite{glasner2003coarsening,glasner2005collision,dai2010mean}.
Further studies have provided rigorous analysis establishing upper bounds on the coarsening rate \cite{otto2006coarsening,dai2010mean} 
and understanding of the dynamics in the full two-dimensional problem \cite{dai2011ostwald,glasner2009ostwald,pismen2004mobility}.
These results follow from studies of coarsening in other systems in materials science having a global conservation of mass \cite{kohn2002upper}, with coarsening proceeding through a sequence of unstable states until reaching the final single-droplet stable steady state. For fluids experiencing evaporation or condensation, there is no conservation of mass, and new approaches are needed to describe the richer dynamics that can occur. 
Physical modeling of thermally-driven dropwise condensation has been a longstanding area of study \cite{rose1973dropwise,rose2002proc} that has gained a strong level of interest in connection with recent engineering applications for improving heat transfer from high-power devices \cite{enright2014dropwise,wen2017hydrophobic,wen2017wetting,anderson2012,ryk2013}
and desalination \cite{hou2015recurrent}. Limitations of experiments can make it difficult to collect enough data on stages of coarsening of droplets to fully resolve the long-time behaviors \cite{boreyko2009self,ryk2013,hou2015recurrent}, hence it is very helpful to explore the predictions that can be obtained from more detailed models for these dynamics.

\par
At low temperatures, fluids may be only very weakly volatile and hence conserve mass to high precision,  but more generally they experience phase change at very slow rates involving a thermodynamic balance between the liquid state and the ambient vapor phase in the surrounding atmosphere. The simplest descriptions for how the volatility of fluids can influence long-time coarsening dynamics use a one-sided model \cite{burelbach1988nonlinear} which treats the vapor phase as passively responding to the dynamics of the thin film. Evaporation or condensation is driven by temperature differences between the fluid layer and the surrounding vapor \cite{wayner1999intermolecular}. We show that the influence of slow condensation can be incorporated into coarsening models of thin films, and will have dramatic effects on the long-time behaviors.

\par 
Burelbach et al. \cite{burelbach1988nonlinear} presented a very influential one-sided lubrication model for volatile thin films that in simplified form can be expressed as
\begin{equation}
{\partial h\over \partial t} = {\partial \over \partial x}\left( h^3
{\partial p \over \partial x}\right) -
\frac{ \beta p}{h+\K}.
\lbl{Mainpde}
\end{equation}
This is an equation for the evolution of the free surface height of the fluid film, $h(x,t)$ in response to transport within the film (the first term on the right side) and loss or gain of mass due to phase change from surrounding vapor (the second, non-conservative flux term). Here $\beta\ge 0$ is effectively a phase change rate and $K>0$ is called a kinetic parameter \cite{Thiele2014,ajaev2005spreading,ajaev2005evolution,kumar2019,maki2011}.
The dynamic pressure of the interface \cite{wayner1999intermolecular} is given by
\begin{equation}
p=\Pi(h)-\pstar- {\partial^2 h\over \partial x^2},
\lbl{defp}
\end{equation}
where $\partial^2h/\partial x^2$ gives the linearized curvature of the free surface 
and $\pstar$ gives the influence of a spatially uniform temperature difference between the liquid layer and the vapor phase \cite{ajaev2005spreading,ji2018instability}. 
The first term, $\Pi(h)$, is a disjoining pressure describing intermolecular forces between the fluid and the solid substrate it coats. For convenience, we use the simple form \cite{oron1999dewetting,oron2001dynamics,schwartz2001dewetting}
\begin{equation}
\Pi(h)= \frac{\eps^2}{h^3}\left(1-\frac{\epsilon}{h}\right),
\lbl{pi}
\end{equation}
describing the influence of van der Waals forces on a hydrophobic substrate, with $h=O(\eps)>0$ defining a strongly bound adsorbed or ``precursor'' layer which sets a lower bound on film thicknesses \cite{bertozzi2001dewetting,ji2018instability}. The maximum disjoining pressure is
$\pmax=27/(256\eps)=\Pi(4\eps/3)$, and the property $\Pi(h\to\infty)\to 0$ denotes that molecular forces are negligible for thick films. 
\par
\begin{figure}
    \centering
\mbox{\includegraphics[height=1.26in]{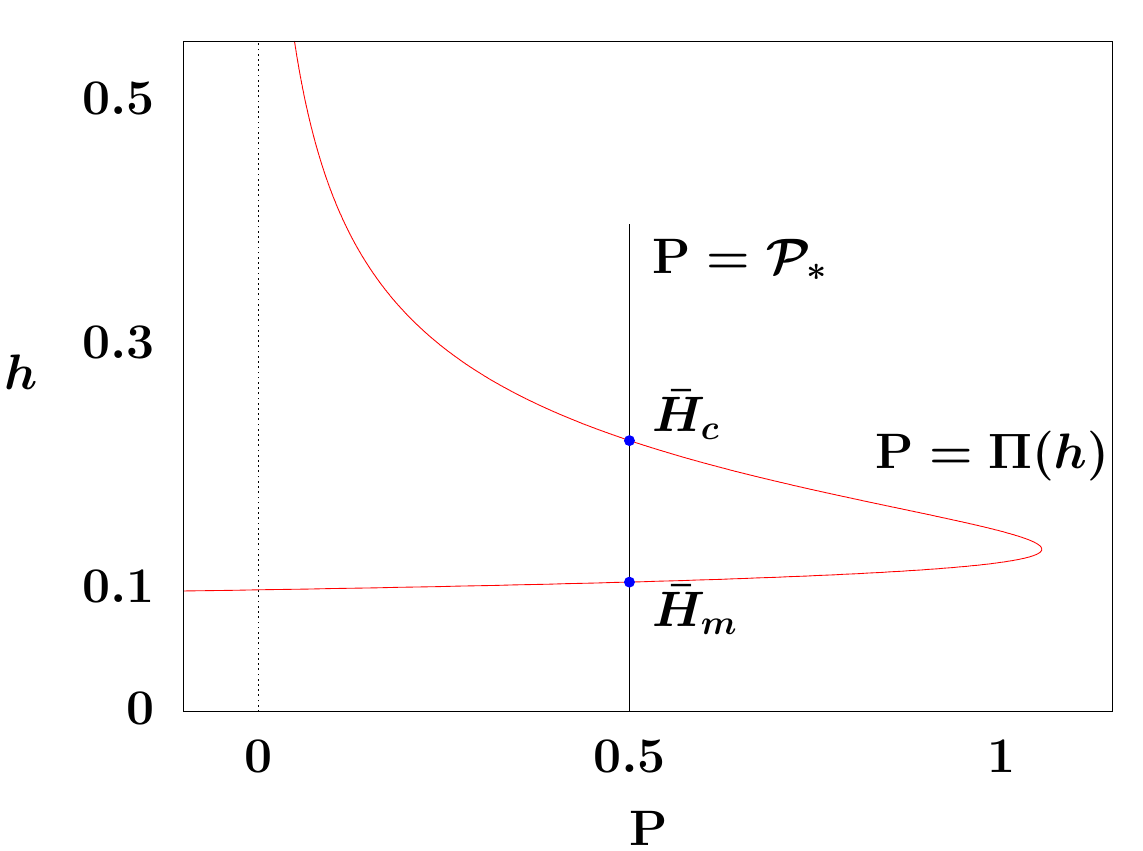}
   \includegraphics[height=1.26in]{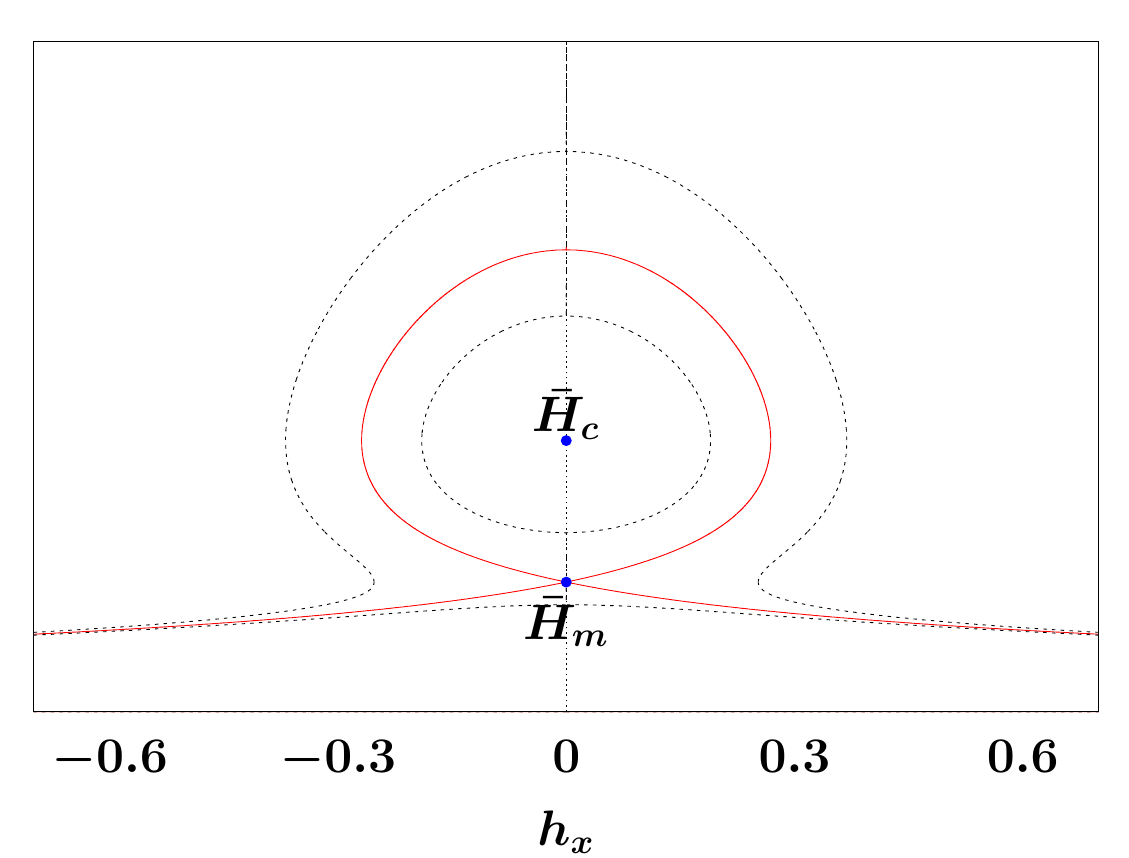}
   \includegraphics[height=1.26in]{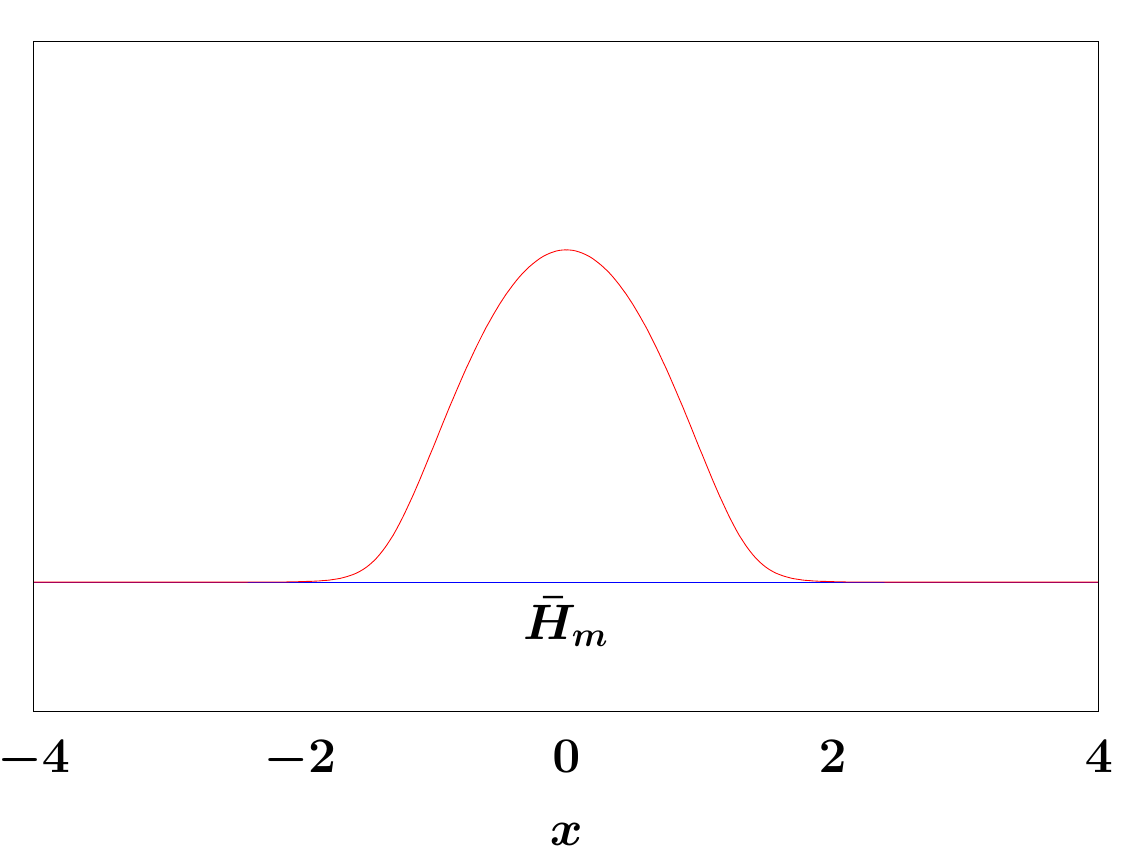}}
    \caption{(Left) The disjoining pressure function $\pps=\Pi(h)$, (middle) orbits in the $h,h_x$ phase plane for steady solutions, (right) the homoclinic steady droplet profile for $p=0$ ($\pps=\pstar$). All three are drawn with the common vertical axis, $0\le h\le 0.5$.}
    \label{fig1}
\end{figure}

This model is a gradient flow in terms of the energy
\begin{equation}
\E[h] = \int_0^L\frac{1}{2}
\left({\partial h\over \partial x}\right)^2 + U(h)\, dx\,,
\lbl{Energy}
\end{equation}
with $U'(h)=\Pi(h)-\pstar$, making this energy monotone dissipated,
\begin{equation}
\frac{d\E}{dt} = -\left(
\int_0^L h^3\left(\frac{\partial p}{\partial x}\right)^2~dx +\beta
\int_0^L\frac{p^2}{h+\K} ~dx\right) \le 0\,.
\lbl{Dissipation}
\end{equation}
Consequently, for general $\beta>0$, equilibria must have a spatially uniform pressure (for the first integral to vanish) and to make the second integral vanish, we need either $p\equiv 0$ or $h\to\infty$ for $t\to\infty$. 
The first case ($p\equiv 0$) corresponds to equilibria like the droplet shown in Fig.~\ref{fig1}, the latter case ($h\to\infty$) describes spatially uniform ``filmwise'' condensation.
For $\pstar>0$ there is no lower bound on the energy \eqref{Energy}, but the long-time attracting state on any finite-time domain can be shown to be a uniformly condensing film satisfying
\begin{equation}
    {dh\over dt} \sim {\beta \pstar\over h} \qquad \implies \qquad h(t)\sim \sqrt{2\beta \pstar t} \qquad\mbox{for
    $t\to\infty$}.
    \lbl{flood}
\end{equation}
For $\beta>0$, 
the total fluid mass is not conserved and its evolution depends strongly on the dynamics of the pressure,
\begin{equation}
\M(t)=\int_0^L h~dx, \qquad \frac{d\M}{dt} = -\beta \int_0^L
\frac{p}{h+\K}~dx.
\lbl{Rateofmass}
\end{equation}
Namely the mass will locally decrease or increase due to local variations of the pressure, $p>0$ or $p<0$, respectively.
\par
The constant $\pstar$ originates from the scaled temperature difference between the fluid-solid interface and the saturation temperature, and its value is crucial for the dynamics of the system.
If $\pstar >\pmax$ then strong condensation dynamics soon leads to \eqref{flood}. If $\pstar<0$ then evaporation dominates, leading to $h\to O(\eps)$ everywhere. 
In the critical range, $0 < \pstar \le \peps$, 
the disjoining pressure can compete with phase change effects yielding complex dynamics. In this range, the problem has two flat-film steady states $\Hm$ and $\Hc$  \cite{ji2018instability}, see Fig~\ref{fig1}(left).
For $\eps\to 0$, the smaller flat-film steady state $\Hm$ defines a stable minimum-thickness adsorbed layer,
\begin{equation}
\Hm = \eps + \pstar\eps^2+4\pstar^2\eps^3 + O(\eps^4),
\end{equation}
and the larger, unstable flat-film steady state $\Hc$ is given by
$\Hc = \pstar^{-1/3}\eps^{2/3}+O(\eps)$, that separates filmwise condensation/evaporation dynamics. Setting $p=0$ in \eqref{defp}, these states are fixed points in the phase plane describing a family of spatially periodic steady solutions (about the $\Hc$ center point) with the largest amplitude solution being a homoclinic orbit through the hyperbolic saddle $\Hm$; this solution uniquely defines a critical steady droplet profile. It is unstable and perturbations can drive it towards either condensing or evaporating dynamics. 
\par
Recognizing the important role of the parameter $\pstar$ in this model, for the rest of this paper it will be convenient to describe unsteady droplets in terms of a shifted-pressure, $\pps\equiv p+\pstar$, so that droplets that are larger (smaller) than the equilibrium drop (with $\pps=\pstar$) have shifted-pressures below (above) $\pstar$ respectively, $0< \pps < \pstar$ ($\pstar<\pps<\pmax$).

\par
We focus on the limit $\beta\to 0$, where the total mass evolves slowly, with rate $O(\beta)$. We believe this may be the simplest regime among many classes of complicated dynamics possible for \eqref{Mainpde}. Considerations for the scaling of the $\beta$ parameter will be addressed at several points in the text -- first for convenience in formal asymptotics in Section~2 and later, further issues on local and larger-scale dynamics will be described.
In Section~2 we extend the approach from \cite{glasner2003coarsening} to reduce the governing PDE to a dynamical system for the evolution of interacting quasi-steady droplets. Sections 3 and 4 examine the behaviors for this system for the low-dimensional cases of a single drop or a pair of drops in periodic domains. This background will then be used to gain some understanding of the problem of dynamics for larger sets of droplets (Section~5) and ultimately the statistical scalings that describe the stages of coarsening that can occur (Section~6).

\section{Derivation of the droplet dynamics model}
We begin by reducing the PDE \eqref{Mainpde} for general solutions to a lower-dimensional dynamical system describing interacting fluid droplets that form after initial transients \cite{glasner2003coarsening,glasner2005collision,kit2011,witelski2020}. 
For $\beta\to 0$, the influence of fluid volatility has a weak influence on the dissipation of energy and in this limit a continuous family of quasi-steady drops exists, parametrized by spatially uniform pressure in $0<\pps<\pmax$. 
\par
Following \cite{glasner2003coarsening}, we begin by constructing a proto-type model for the dynamics of a single near-equilibrium droplet $h=h(x,t)$ subject to surrounding influences. Consider a droplet starting in the center of a finite domain with some initial pressure,
$$
h(x,0)=\Hd(x;\pps_0)\qquad -\ell\le x \le \ell
$$
and 
mass fluxes, defined as 
\begin{equation}
J\equiv -h^3 \partial_x p,
\end{equation}
are imposed with fixed values at the domain boundaries,
$$
J(-\ell) = \sigma \tilde{J}_-\qquad J(\ell) =\sigma \tilde{J}_+
$$
with $\sigma\ll 1$  and $\tilde{J}=O(1)$ describing weak fluxes so the evolution of the droplet is slow and can be assumed to be quasi-steady.
We introduce a time scale  $\tau = \sigma t$,
where $\sigma \ll 1$ will be selected by the scale of the non-conservative flux $\beta$.
By assuming that the
evaporation/condensation effects and the fluxes cause the droplet to vary
slowly in time, we write an expansion perturbing a leading order quasi-static droplet, 
\begin{equation}
    h(x,t) = \Hd(x-\X(\tau); \pps(\tau))+\sigma h_1(x,\tau)+O(\sigma^2),
    \lbl{Eq:coarsen_perb}
\end{equation}
where the droplet profile $\Hd(x;\pps)$ satisfies
\begin{equation}
    \Pi(\Hd)-{d^2\Hd\over dx^2} = \pps\,,
\lbl{eq:quasi-static}
\end{equation}
this is a homoclinic solution that approaches a far-field flat film $\Hd(|x|\to\infty)\sim \eps +\eps^2 \pps \sim \Hm +O(\eps^2)$.
Substituting \eqref{Eq:coarsen_perb} into \eqref{Mainpde} and using \eqref{eq:quasi-static}
yields the leading-order equation 
\begin{equation}
    \sigma\left[-\frac{d \Hd}{d x}\frac{d\X}{d\tau} +
\frac{\partial\Hd}{\partial \pps}\frac{d\pps}{d\tau}\right] =
-\frac{\beta (\pps-\pstar)}{\Hd+\K}+\sigma \mathcal{L}h_1 ,
    \lbl{Eq:leading-order}
\end{equation}
where the linear operator $\mathcal{L}$ is given by the composition of two
second-order operators, $\mathcal{L}\Psi=\mathcal{RS}\Psi$ \cite{ji2018instability}. 
The operator $\mathcal{R}$ includes the evaporative flux and the
mobility function, and the operator $\mathcal{S}$ is the linearized pressure
operator,
\begin{equation}
    \mathcal{R}w \equiv -\frac{\beta w}{\Hd+\K}+\frac{\partial}{\partial
x}\left(\Hd^3\frac{\partial w}{\partial x}\right), \quad
\mathcal{S}v\equiv \Pi'(\Hd)v - \frac{\partial^2 v}{\partial x^2}.
\end{equation}
The adjoint operator of $\mathcal{L}$ is given by
$  \mathcal{L}^{\dag} \Phi\equiv \mathcal{S}\mathcal{R}\Phi$.
Expanding $\mathcal{L}$
for $\beta \to 0$, 
$\mathcal{L}\Psi = \mathcal{L}_0\Psi + \beta \mathcal{L}_1\Psi +
O(\beta^2)$,
gives the leading order operator as the
mass-conserving 
operator,
\begin{equation}
    \mathcal{L}_0\Psi = \mathcal{R}_0 \mathcal{S}\Psi, \quad \mathcal{R}_0 w
\equiv \frac{\partial}{\partial x}\left(\Hd^3\frac{\partial w}{\partial
x}\right).
\end{equation}
and the adjoint operator 
is $\mathcal{L}_0^{\dag} \Phi\equiv \mathcal{S}\mathcal{R}_0\Phi$.
\par
To address the influence of condensation effects on the droplet dynamics, we select the timescale $\sigma\sim\beta\ll 1$ so that the non-conservative term is included in the leading order dynamic balance \eqref{Eq:leading-order}, with $\sigma\ll 1$ being necessary to ensure that the dynamics are slowly evolving. We choose to study the regime $\beta\ll 1$ so that the quasi-steady droplets are determined by \eqref{eq:quasi-static} and we can take advantage of results from previous studies \cite{glasner2003coarsening,glasner2005collision}. 
\par
Based on the spatial fluxes between finite-sized near-equilibrium drops it can be shown that $\sigma=O(\eps^3)$ \cite{glasner2003coarsening}.
Consequently, we define a reduced evaporation/condensation rate $\ba$ as $\beta=\ba\eps^3$ for the following formal asymptotics, with $\ba=O(1)$. Further dynamic considerations described in later sections will lead us to define the weak condensation regime in terms of focusing on small $\beta_0\to 0$. In Section~\ref{sec:singleDrop} we show that using $\beta_0\gg O(1)$ changes some of the properties of the quasi-steady droplets and requires a different line of analysis.  
Then the leading-order dynamic equation \eqref{Eq:leading-order} at $O(\beta)$ reduces
to
\begin{equation}
    -\frac{d \Hd}{d x}\frac{d\X}{d\tau} +
\frac{\partial\Hd}{\partial \pps}\frac{d\pps}{d\tau} = -\frac{\ba
(\pps-\pstar)}{\Hd+\K}+\mathcal{L}_0 h_1.
\lbl{linearizedPDE}
\end{equation}
The null space of $\mathcal{L}^{\dag}_0$ is spanned by the two bounded
functions \cite{glasner2003coarsening},
\begin{equation}
    \Psi_1(x) = 1, \quad \Psi_2(x) =
\int_0^x\frac{\Hd(x')-\Hm}{\Hd(x')^3}\,dx'.
    \lbl{Eq:reduced_leading-order}
\end{equation}
Taking the inner product of \eqref{linearizedPDE} with $\Psi_1(x-\X)$ yields the equation for the evolution of the total mass, including the contribution of the non-conservative flux,
\begin{equation} 
{d\mathcal{M}\over d\tau} = -\tilde{J}_++\tilde{J}_- -\beta_0 \Jncl\qquad
\Jncl= \int_{-\ell}^\ell \frac{\pps-\pstar}{\Hd+\K}~dx .
\lbl{Jnc29}
\end{equation}
If the droplet is located away from the edges of the domain, then small changes in $\X$ have a negligible effect on the mass, so $\mathcal{M}=\int \Hd(x-\X;\pps)\,dx\approx \mathcal{M}(\pps)$. Using the chain rule we can re-write this as
\begin{equation}
    \frac{d\pps}{dt} =C_P(\pps)\left[J_+-J_-
    +\beta \Jncl\right],
    \qquad C_P(\pps)=  - \left({d\mathcal{M}\over d\pps}\right)^{-1}\,. 
    \lbl{Eq:dpdt_scaled}
\end{equation}
Similarly, using the fact that $\Hd$ has a symmetric profile and therefore $\Hd$ is odd with respect to the center of the drop, taking the inner product of $\Psi_2(x-\X)$ with
\eqref{linearizedPDE} yields an equation for the evolution of the center of mass,
\begin{equation}
    \frac{d\X}{dt} = -C_X(\pps) \left[{J}_++{J}_-\right],
\qquad 
C_X(\pps)= \frac{1}{2}\int_{-\ell}^\ell
\frac{\Hd-\Hm}{\Hd^3}\,dx\bigg/ \int_{-\ell}^\ell
\frac{(\Hd-\Hm)^2}{\Hd^3}\,dx\, .
\lbl{Eq:dxdt_scaled}
\end{equation}
Relations to define the fluxes and evaluate the integrals must still be specified, but
for $\beta=0$ equations (\ref{Eq:dpdt_scaled}, \ref{Eq:dxdt_scaled}) have the same form as the model
derived in \cite{glasner2003coarsening}. The same form of the quasi-steady droplets is being used, however, we will see that the non-conservative effects will dramatically change the fluxes between drops.

\subsection{The droplet core approximation and the non-conservative flux}

For a quasi-static droplet that satisfies \eqref{eq:quasi-static}, with the pressure in the range $0< \pps<\peps$,
in the core region of the droplet,  $\Hd\gg\eps$, and the disjoining pressure $\Pi(\Hd)$ is negligible. Therefore, using the balance of the surface tension $\Hd_{xx}$ and the pressure $\pps$, we  approximate the core of the droplet by a parabola \cite{bertozzi2001dewetting,glasner2003coarsening,kit2012,otto2006coarsening}, 
\begin{equation}
    \Hd(x;\pps)\approx \frac{1}{2}\pps(w^2-x^2)\quad \mbox{for } |x|\ll w,
\lbl{eq:drop_parabolic}
\end{equation}
with $w$ being the half-width or radius of the droplet,
and $\Hd\approx \Hm$ elsewhere. 
Droplets are homoclinic solutions asymptoting to $\Hm$ in the far-field; separating the compactly-supported parabolic core approximate profile from the surrounding adsorbed precursor film is a convenient and widely used approach for the analysis of this system \cite{kit2011,otto2006coarsening}.
Fig.~\ref{fig:schematic} (right) shows two typical quasi-static droplets and the parabolic profiles approximating their core regions.
\par
From the first integral of \eqref{eq:quasi-static} and the estimate that the inter-droplet film $\Hd\approx \Hm$, the maximum height of the droplet $h_{\max}$ satisfies $U(h_{\max}) - U(\Hm) = \pps (h_{\max}-\Hm)$. In the limit $\eps\to 0$, we get the leading-order estimate $h_{\max}\sim -U(\Hm)/\pps$. In this limit, the half-width of the droplet is given by \cite{glasner2003coarsening,kit2012}
\begin{equation}
w(\pps)\sim A/\pps\qquad \mbox{with}\quad A = \sqrt{2|U(\Hm)|},
\lbl{eq:drop_width}
\end{equation}
and for \eqref{pi} this yields $A=1/\sqrt{3}$ to leading order for $\epsilon \to 0$.
The mass of the droplet then can be approximated by the mass of the core region as
\begin{equation}
{M}(\pps) = \int_{-w}^{w} \Hd(x;\pps)~dx={2\over 3} \pps w^3 \sim \frac{2A^3}{3\pps^2},
\lbl{eq:drop_mass}
\end{equation}
where the last relation uses \eqref{eq:drop_width}.

Estimates of the pressure coefficient function $C_P(\pps)$ (from \eqref{Eq:dpdt_scaled} with $\mathcal{M}\approx M$) and the drift coefficient function $C_X(\pps)$ were found in \cite{glasner2005collision} as
\begin{equation}
\lbl{eq:coeff}
C_P(\pps) \sim \frac{3\pps^3}{4A^3} > 0,\qquad C_X(\pps)\sim \frac{B}{\eps\ln(\pps/\peps)} > 0,
\end{equation}
where $B$ is a negative constant independent of $\eps$.  
\par
In order to complete the descriptions of \eqref{Eq:dpdt_scaled} and \eqref{Eq:dxdt_scaled} we must express the fluxes in terms of droplet properties.
The non-conservative flux \eqref{Jnc29} is defined on the entire domain but can be separated into  two parts. We first focus on its contributions in the droplet core region and will account for the impact of this flux on the surrounding thin films separately in the next section.
\par
Using approximation \eqref{eq:drop_parabolic}, the non-conservative flux in the droplet core region is
\begin{equation}
    \Jnc 
    \approx\int_{-w}^{w}
\frac{\pps-\pstar}{\textstyle\frac{1}{2}\pps(w^2-x^2)+K}~dx.
\lbl{eq:non-cons_flux}
\end{equation}
This integral relates to the loss or gain of mass in the droplet core. The remainder of the $\Jncl$ flux (on $|x|> w$) will be handled implicitly by calculating the pressure in the precursor films between drops.

\par
Integral \eqref{eq:non-cons_flux} can be evaluated to yield the estimate
\begin{equation}
\Jnc(\pps)\approx {4\over A}(\pps-\pstar)\bb(\pps)
{\mbox{arctanh}[\bb(\pps)]}\quad \mbox{with}
\quad \bb(\pps) =\left[1+{2K\pps\over A^2} \right]^{-1/2}.
\lbl{eq:non_cons_flux}
\end{equation}
This form makes clear that the sign of the non-conservative flux $\Jnc$ is determined by the relation between $\pps$ and $\pstar$. For an isolated quasi-static droplet with pressure $\pps > \pstar$, 
the non-conservative flux will be positive, increasing the pressure and decreasing droplet mass due to evaporation.  If the droplet pressure is smaller than the critical pressure, $0 < \pps < \pstar$, the flux will be negative and the pressure will decrease, corresponding to increasing droplet mass due to condensation.


\subsection{Mass fluxes between drops}
\lbl{sec22}
\begin{figure}
    \centering
    \includegraphics[height=4cm]{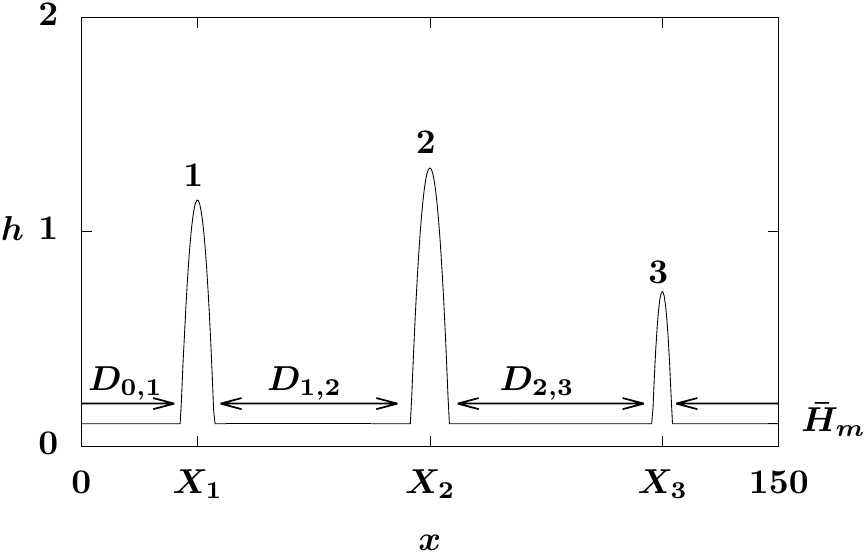}\quad 
    \includegraphics[height=4.1cm]{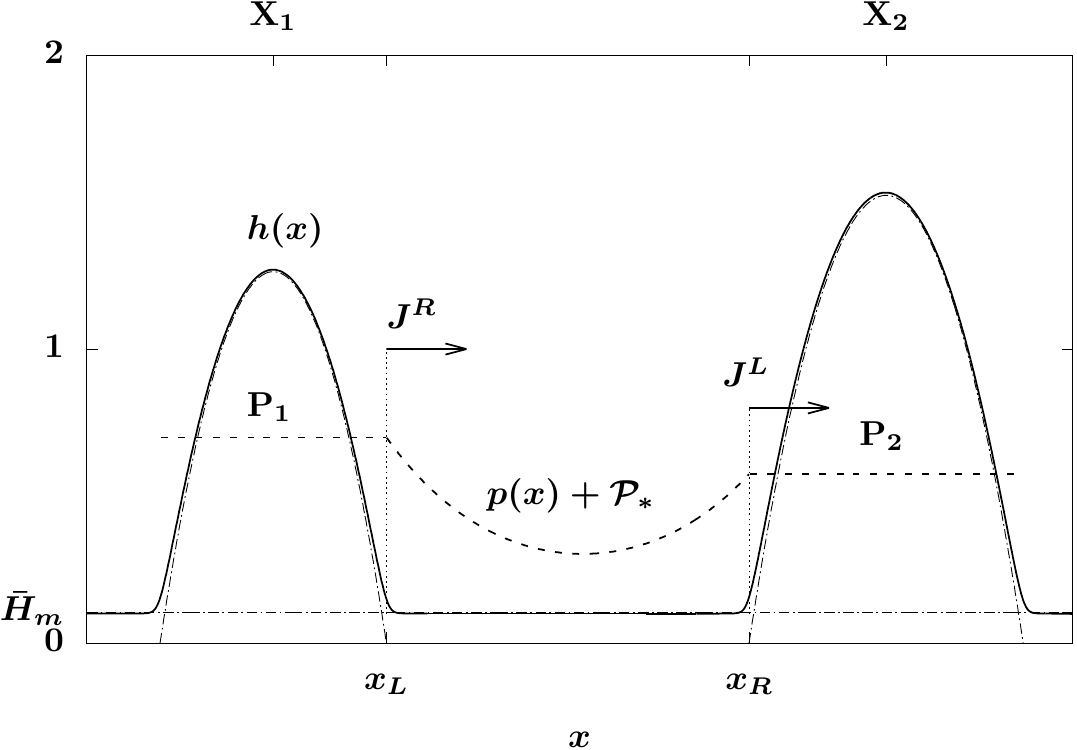}
    \caption{(Left) A schematic figure of a system of three quasi-static droplets in a periodic domain $0 \le x \le 150$. (Right) Two quasi-static droplets from a numerical solution of \eqref{Mainpde} (solid curve) locally satisfying \eqref{eq:quasi-static} 
    with the pressures $\pps=\pps_{1,2}$, 
    where the core of droplets can be approximated by parabolas (dot-dashed curves) defined by \eqref{eq:drop_parabolic}. The shifted pressure $\pps=p(x)+\pstar$ (dashed curve) for the precursor layer is given by \eqref{eq:pressure_precursor} over $x_L\le x\le x_R$ where the fluxes at the edges of droplets are $J^{R,L}$. 
    }
    \lbl{fig:schematic}
\end{figure}
\par 
To model the fluxes between neighboring droplets, we focus on the thin film between the drops.
Assume that the film between two drops is thin, $h\approx \Hm$,
and quasi-steady. Then the reduced governing equation \eqref{Mainpde} between drops,
\begin{equation}
     {d\over dx}\left(h^3 {dp\over dx}\right) - {\beta p\over h+\K}=0,
\end{equation}
can be approximated by
\begin{equation}
 \Hm^3 {d^2p\over dx^2} - {\beta\over \Hm+\K}\, p=0\qquad \mbox{on $x_L\le x\le x_R$},
 \end{equation} 
 subject to boundary conditions on the (unshifted) pressures of the adjacent drops, $p(x_L)=p_L$ and $p(x_R)=p_R$.
 This linear problem has the general solution
\begin{equation} 
p(x)=A\cosh(\alpha x) + B\sinh(\alpha x), \qquad \alpha=\sqrt{\beta \mu/\Hm^{3}}\qquad 
\mu={1\over  \Hm+\K}
\end{equation}
and applying the boundary conditions yields
\begin{equation}
 p(x)= {p_R\sinh(\alpha[x-x_L]) +p_L\sinh(\alpha[x_R-x])
\over \sinh(\alpha D)}\qquad \mbox{on $x_L \le x \le x_R$}\, ,
\label{eq:pressure_precursor}
\end{equation}
where $D=x_R-x_L$ is the distance between contact lines of adjacent droplets; see Fig.~\ref{fig:schematic} in terms of droplets 1 and 2, with $p_L=\pps_1-\pstar, p_R=\pps_2-\pstar$ and
$x_L=\X_1+w(\pps_1), x_R=\X_2-w(\pps_2)$.
\par
For $\alpha\to 0$ (i.e.\ $\beta\to 0$) this reduces to a linear pressure profile,
$p=\GLR(x-x_L)+p_L$ with pressure gradient $\GLR=(p_R-p_L)/D$, and a uniform flux between the drops,
$J = -h^3\partial_x p\approx -\Hm^3 \GLR$ as previously used in coarsening models for non-volatile films \cite{glasner2003coarsening,kit2011}.
For $\beta>0$, the fluxes at the neighboring drops will differ and must be distinguished. The flux $J^R$ at the right edge of a drop with pressure $p_L$ whose contact line  at $x_L$ is the left end of a precursor film connecting it to a neighboring droplet with pressure $p_R$ and contact line at $x_R$ (see Fig.~\ref{fig:schematic}(Right)),
{\small
\begin{equation}
\lbl{hmflux}
J^R(x_L)= -{\Hm^3 \alpha (p_R-p_L\cosh(\alpha D)) \over
\sinh(\alpha D)}
\sim 
-\Hm^3\GLR + {\beta\mu\over 6} D(2p_L+p_R) +O(\beta^2).
\end{equation}
}
\par\noindent
and similarly the flux $J^L$ at the left edge of a drop whose contact line at $x_R$ is the right boundary of the  precursor film is
{\small
\begin{equation} 
\lbl{hmflux2}
J^L(x_R)= -{\Hm^3  \alpha (p_R\cosh(\alpha D) -p_L)\over
\sinh(\alpha D)}
\sim  -\Hm^3\GLR - {\beta\mu\over 6} D(p_L+2p_R)+O(\beta^2).
\end{equation} 
}
\par\noindent
The derivation of these fluxes has focused on the thin film region between droplets in this section, but in general, our focus will be on using these fluxes to describe the coupling between the adjacent droplets. In particular, we see that these fluxes include both mass-conserving exchanges between drops (the $\GLR$ terms) and non-conservative effects due to phase change (the $O(\beta)$ terms). To separate out these effects, we have assumed that $\alpha D\ll 1$. Using $\Hm\sim \eps$, this condition can be re-written as $\beta\ll \K\eps^3/D^2$; this is more restrictive than the original assumption used to obtain \eqref{linearizedPDE}, but is very helpful for interpreting behaviors, so we will restrict ourselves to this regime to define weak condensation.

\subsection{The simplified dynamical model}
\label{sec:simplifiedDynamicalModel}
We can now describe an array of $N$ slowly-varying droplets parameterized by their positions $\{\X_k(t)\}$ and pressures $\{\pps_k(t)\}$, where $k=1,2,\cdots, N$ (see Fig.~\ref{fig:schematic}). The governing equations that describe the evolution of the coupled $\{(\X_k, \pps_k)\}$ are given by
\begin{subequations}
\lbl{eq:ODE_system}
\begin{eqnarray}
    \frac{d\pps_k}{dt}& =& C_P(\pps_k)(J_{k,k+1}^R-J_{k-1,k}^L+\beta \Jnc_k),
\lbl{dPk}\\[4pt]
\frac{d\X_k}{dt} &=& -C_X(\pps_k)(J_{k,k+1}^R+J_{k-1,k}^L), 
\lbl{dXk}
\end{eqnarray}
\end{subequations}
where for $\beta\to 0$ the leading-order mass fluxes between neighboring droplets are
\begin{subequations} 
\begin{eqnarray}
    J_{k,k+1}^R &=  & -\Hm^3\frac{\pps_{k+1}-\pps_k}{D_{k,k+1}}+\frac{\beta\mu}{6}D_{k,k+1}(2\pps_{k}+\pps_{k+1}-3\pstar), 
    \lbl{JRkkp}\\
    J_{k-1,k}^L &= & -\Hm^3\frac{\pps_{k}-\pps_{k-1}}{D_{k-1,k}}-\frac{\beta\mu}{6}D_{k-1,k}(\pps_{k-1}+2\pps_{k}-3\pstar),
    \lbl{JLkmk}
\end{eqnarray}
where the length of the film between two neighboring droplets (the distance between their contact lines) is (see Fig.~\ref{fig:schematic})
\begin{equation}
    D_{k,k+1} = (\X_{k+1}-w(\pps_{k+1})) - (\X_{k}+w(\pps_k)),
    \lbl{dkkp}
\end{equation}
and the non-conservative flux \eqref{eq:non_cons_flux} can be written as
\begin{equation}
 \lbl{eq:noncons_flux}
    \Jnc_k\sim
    \begin{cases}
    {2\pstar\over A}\ln\left({K\pps_k\over 2A^2}\right) & \pps_k\to 0,\\[4pt]
    {2A\over K} & \pps_k \gg \pstar .
    \end{cases}
   \end{equation} 
\end{subequations}
System \eqref{eq:ODE_system} describes smooth, quasi-steady evolution of droplets until one of two possible singular behaviors  (called \textit{coarsening events}) arise generating fast dynamics that decreases the number of droplets:

\begin{enumerate}
\item[(a)]  \emph{Droplet collapse}:  Quasi-steady droplets exist for the range of 
pressures, $0< \pps_k <\pmax$.  
The limit $\pps\nearrow \pmax$
corresponds
to small droplets effectively vanishing into the $\Hm$ adsorbed film.
 It was shown that this collapse happens in finite time with $\pps= O((T_c-t)^{-1/3})$ \cite{glasner2003coarsening}. 
     In practice, we define a droplet collapse event to occur when the droplet pressure reaches a fixed critical pressure:
    \begin{equation}
        \mbox{Droplet $k$ collapses if $\pps_k(t)=\pmaxh$},
    \lbl{eq:collapseCondition}
    \end{equation}
    with $\pmaxh:= (1-\eta)\peps$
    where $\eta>0$ is a small parameter; $\pmaxh$ defines a scale for the minimum observable quasi-stable droplet.
    
    At a collapse event, droplet $k$ is removed to yield the reduced system with $N-1$ drops for further evolution in \eqref{eq:ODE_system}.

    \item[(b)] \emph{Droplet collision}: 
        When the separation distance between the edges of adjacent drops vanishes, after a rapid transition, the two droplets will merge to produce a single combined drop. 
    This has previously been called droplet collision and occurs when the flux \eqref{JRkkp} between drops diverges with $D_{k,k+1}(t)\searrow 0$.
   \par
    In practice, we define a droplet collision event to occur when the distance between a pair of  drops \eqref{dkkp} reaches a fixed small distance, $\Dmin>0$:
    \begin{equation}
        \mbox{Droplets $k$ and $k+1$ collide if $D_{k,k+1}(t) = \Dmin$.}
    \lbl{eq:collision_criteria}
    \end{equation}
Having a positive $\Dmin$ gives a heuristic correction to the parabolic approximation for droplets \eqref{eq:drop_parabolic} under-estimating effective droplet widths.
\par

     When a droplet collision event occurs, we follow  \cite{glasner2005collision} and assume that merging occurs rapidly so that evaporation or condensation are negligible and it is reasonable to conserve the total mass of the two droplets. We also assume that the position of the merged droplet is symmetric with respect to the outer contact lines of the two merging droplets. The pressure and position of the merged droplet $(\pps_{k,k+1},X_{k,k+1})$ after collision are therefore given by
\begin{equation}
    \pps_{k,k+1}  = \left(\frac{1}{\pps_k^2}+\frac{1}{\pps_{k+1}^2}\right)^{-1/2}, \qquad   X_{k,k+1} = \tfrac{1}{2}[\X_k-w_k + \X_{k+1}+w_{k+1}].
    \lbl{postcollide}
\end{equation}
The merged droplet will then be used to replace one of the colliding drops,
$(\pps_{k,k+1},X_{k,k+1})\to (\pps_k, \X_k)$,
and the system \eqref{eq:ODE_system} will restart with $N-1$ droplets.
    
\end{enumerate}

Both mechanisms cause the number of droplets to decrease from $N$ to $N-1$ in the system.
To numerically capture the coarsening event, we halt the coupled ODE system \eqref{eq:ODE_system} when a collapse or collision event is detected. Then the reduced ODE system  \eqref{eq:ODE_system} is restarted for the remaining  $N-1$ droplets in this coarsening-type piecewise-defined dynamical system \cite{watson,budd}. We observed that the dynamics are not sensitive to the regularizations introduced by the parameters $\pmaxh,\Dmin$.
For $\beta = 0$, the dynamical model \eqref{eq:ODE_system} reduces to the coupled ODEs describing the droplet dynamics of the mass-conserving thin film equation \cite{glasner2005collision,glasner2003coarsening}.

\par
To explore the droplet dynamics governed by the dynamical model \eqref{eq:ODE_system}--\eqref{postcollide}, we will consider three droplet configurations on  periodic domains: the dynamics of a single droplet (Sec.~\ref{sec:singleDrop}), the interactions between two droplets (Sec.~\ref{sec:twoDrop}), and a system with multiple droplets (Sec.~\ref{sec:many_drop}). The single droplet case illustrates the influences of non-conservative effects on periodic droplet arrays. The two-droplet case provides insights into the interplay between inter-drop mass fluxes and non-conservative fluxes by separating the evolution of droplet pressures from their motions. 
The many-droplet configuration extends the pairwise droplet analysis in the two-droplet case to a system of slowly-varying well-separated droplets.
\par
For all the numerical studies in this paper, we set $\eps=0.1$ with $\pstar = 0.5$ in the middle of the critical pressure range and $\K=0.1$ unless otherwise specified \cite{ji2018instability,maki2011}. 
We conduct the PDE simulations for the model \eqref{Mainpde} using centered finite differences with backward Euler time stepping. The simplified dynamical system \eqref{eq:ODE_system} is numerically solved using by a Cash-Karp Runge-Kutta scheme with error control \cite{NRC}. Adaptive time stepping is used for both PDE and ODE simulations.

\section{Dynamics of a single droplet: pure condensation}
\lbl{sec:singleDrop}

First, we consider a single droplet $(N=1)$ with the position $\X$ and pressure $\pps$ in a periodic domain $0 \le x \le L$. In this case, the left and right neighbors of the droplet are identical, and (\ref{JRkkp}, \ref{JLkmk}) reduce to the leading-order symmetric inter-drop fluxes
\begin{equation}
    J^R = -J^L \sim \frac{\beta \mu}{2} D(\pps-\pstar),
\lbl{eq:singleDrop_flux}
\end{equation}
where the length of the inter-drop film is
\begin{equation}
    D = L - 2w = L - \frac{2A}{\pps}\, .
\lbl{eq:singleDrop_D}
\end{equation}
Setting $D=0$ yields a minimum pressure, $P_L=2A/L$ corresponding to the largest droplet that can fit in the domain.
From the equal opposing fluxes, \eqref{eq:singleDrop_flux}, the right-hand side of the $d \X/dt$ equation \eqref{dXk} becomes zero. Hence the droplet does not move and the dynamical system \eqref{eq:ODE_system} reduces to
a single equation for the pressure 
\begin{equation}
\lbl{Eqn:dpdt_singleCondense}
    \frac{d\pps}{dt} = C_P(\pps)\left[2J^R+\beta
\Jnc \right].
\end{equation}
Substituting into \eqref{Eqn:dpdt_singleCondense} the forms of the pressure coefficient  $C_P(\pps)$, \eqref{eq:coeff}, the non-conservative flux $\Jnc$ from the core of the droplet, \eqref{eq:noncons_flux}, and the inter-drop fluxes, 
\eqref{eq:singleDrop_flux}, one arrives at the ODE for 
the pressure of the droplet
\begin{equation}
\lbl{Dpdt_singleDropCondense}
\frac{d\pps}{dt} = \frac{3\beta(\pps-\pstar)\pps^3}{4A^3}\left[\mu \left( L-{2A\over \pps}\right)+
{4 \over A} \bb(\pps){\mbox{arctanh}[\bb(\pps)]}
\right].
\end{equation}
This is called the pure condensation case since there are no conservative fluxes between identical droplets ($\gamma=0$) and hence all of the terms on the right are scaled by $\beta$.
\par
For the case $\pps > \pstar$, equation \eqref{Dpdt_singleDropCondense} describes the dynamics of an evaporating droplet that collapses in finite time. In the collapse limit $\pps \gg \pstar$, the leading-order equation of \eqref{Dpdt_singleDropCondense} becomes
\begin{equation}
   \frac{d\pps}{dt}\sim \frac{3\beta \mu L}{4A^3}\pps^4,
\lbl{eq:collapse_evap_limit}
\end{equation}
which yields
$
    \pps(t)\sim \left[\kappa(T_c-t)\right]^{-1/3},
$
where $\kappa = (9\beta\mu L)/(4A^3)$, $T_c$ is the critical blow-up time. 
As $t\to T_c$, the droplet pressure would blow up, $\pps \to \infty$, but in the PDE the pressure is bounded and the droplet collapses to the equilibrium flat absorbed layer $h\equiv \Hm$ due to evaporation based on the droplet collapse criteria \eqref{eq:collapseCondition}.

\begin{figure}
    \centering
    \mbox{
    (a)\includegraphics[height=5cm]{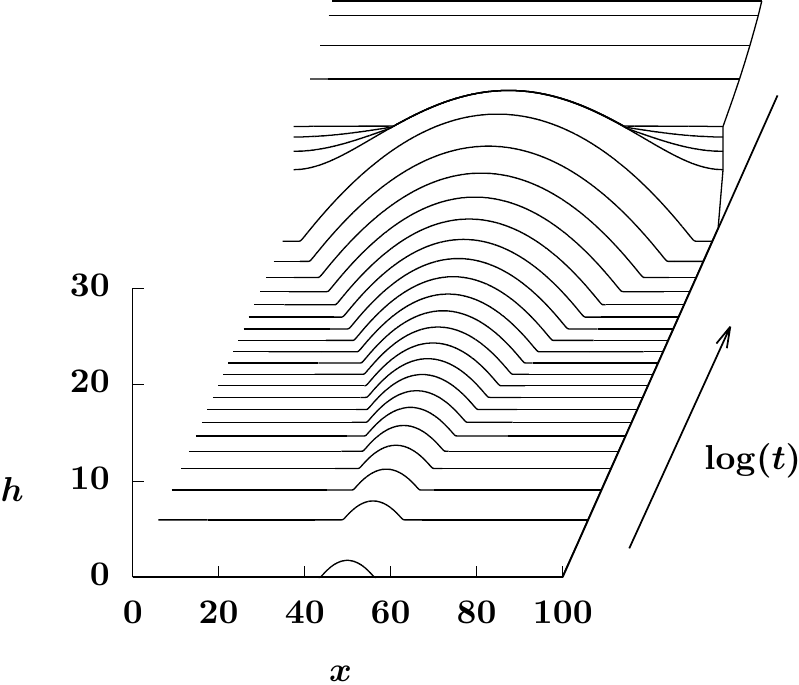}
    (b)\includegraphics[height=4cm]{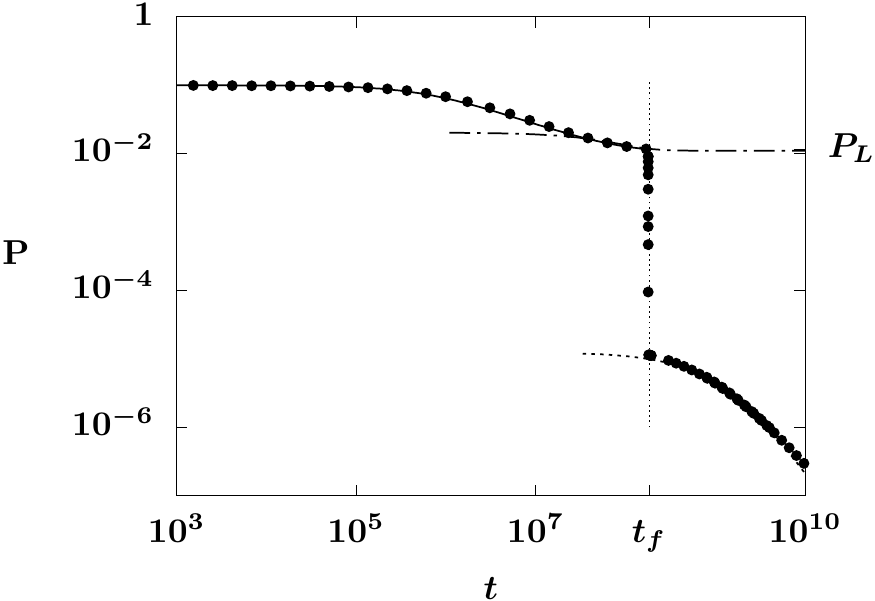}
}
\mbox{} \\
\mbox{
    \hspace{-0.1in}
   (c)\includegraphics[height=4.1cm]{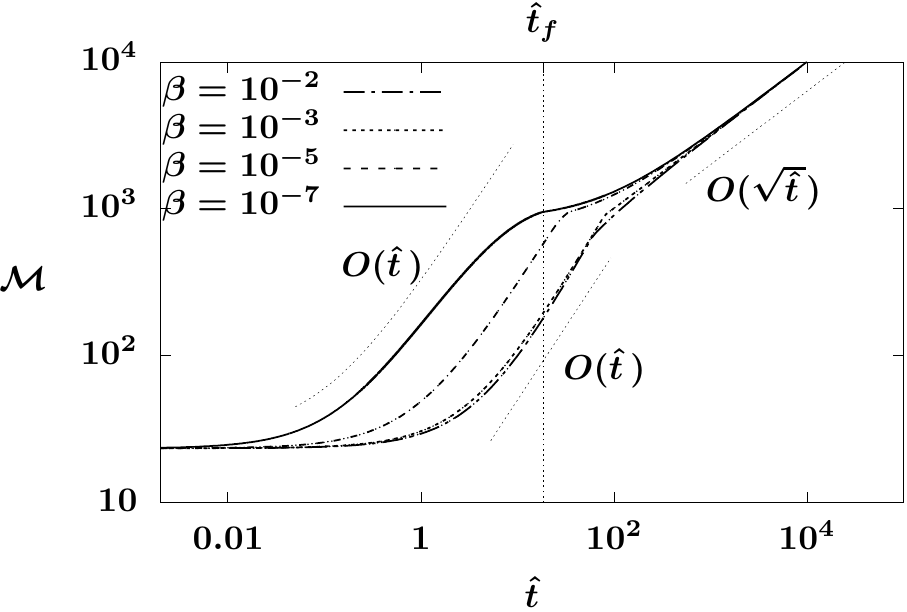}
   (d)\includegraphics[height=3.8cm]{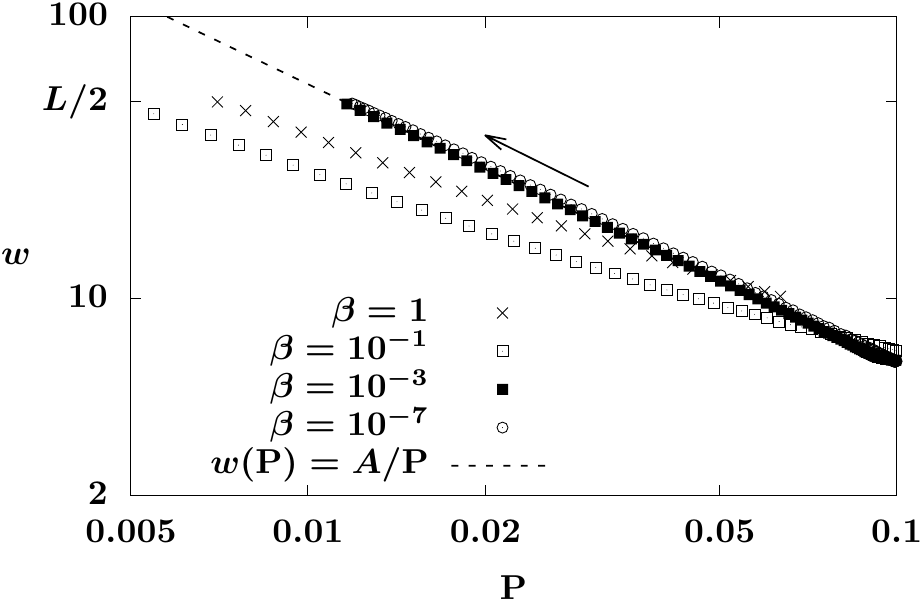}
}
    \caption{(a) PDE simulation of \eqref{Mainpde} for a slowly condensing droplet 
    in a periodic domain $0\le x \le L$ with $L=100$ and $\beta=10^{-7}$. (b) The evolution of the droplet pressure in the PDE model (dots) compared against predictions from the dynamical model \eqref{Dpdt_singleDropCondense} (solid curve) for $t<t_f\approx 1.8\times 10^8$. 
    The last stage of the droplet growth is given by \eqref{eq:condensationLimit} (dot-dashed curve) as $\pps \to \pinf$, followed by convergence to film-wise condensation after the domain is filled for  $t>t_f$ (dotted curve).
    (c) The evolution of the mass $\M(\hat{t}\,)$ for a rescaled time variable $\hat{t}=\beta t$ at several values of $\beta$, showing a transition from the scaling $\M = O(\hat{t}\,)$ in the dropwise condensation stage to the scaling $\M = O(\sqrt{\hat{t}\,})$ in the filmwise condensation stage.  
    (d) Parametric plot of droplet half-width $w(t)$ and pressure $\pps(t)$ from PDE simulations like (a) at several values of $\beta$. The quasi-steady relation $w(\pps)=A/\pps$ \eqref{eq:drop_width} is shown for comparison (dashed line).}
       \label{fig:SingleDropCondense}
\end{figure}

\par
Returning to our primary focus, for the case $\pps < \pstar$,  equation \eqref{Dpdt_singleDropCondense} describes the evolution of a droplet subject to condensation.
Figure \ref{fig:SingleDropCondense}a presents a typical simulation of the PDE \eqref{Mainpde} with $\beta=10^{-7}$ for a slowly condensing droplet placed at the center of a periodic domain $0 \le x \le 100$ (c.f. \cite[Figure 6]{sharma1998equilibrium}). The initial condition is the quasi-static droplet profile $h(x,0)=\Hd(x-\X(0);\pps(0))$ with
$\pps(0)=0.1 < \pstar$ and $\X(0)=50$ at time $t = 0$. 
Figure \ref{fig:SingleDropCondense}b shows that the evolution of the pressure $\pps$ obtained from the PDE simulation (marked by dots) agrees well with the prediction from the dynamical model \eqref{Dpdt_singleDropCondense} (solid curve). This comparison shows that the leading-order linear approximation for the inter-drop fluxes \eqref{eq:singleDrop_flux} provides a good prediction for the long-time droplet dynamics under weak condensation. 

\par
Both terms in the square brackets on the right hand side of \eqref{Dpdt_singleDropCondense} contribute to the evolution of the droplet pressure. The first term is condensation on the film between drops while the second describes condensation in the drop core region. This evolution goes through several stages as $\pps\searrow P_L$.
The rate of condensation is initially dominated by the contributions from the long films between drops, $D\gg 1$. Later, the rate slows as the drops grow and the length of the inter-drop films shrink, $D\searrow \Dmin$.
For a range of intermediate times, when both terms contribute at leading order, the evolution can be approximated by
\begin{equation}
    {d\pps\over dt} \propto - {3\beta\pstar\over 4A^3} \pps^3 \qquad \implies \qquad 
    \pps=O([t-t_0]^{-1/2}) .
    \lbl{new36}
\end{equation}
Using \eqref{new36} and \eqref{eq:drop_mass} we can estimate the growth of the droplet mass as $M=O(t-t_0)$, as shown in Figure~\ref{fig:SingleDropCondense}c, with $t_0$ being a time-shift appropriate for matching to previous stages.

\par
In the final stage, the condensing drop fills almost the entire domain, with $D\to 0$. 
In this limit,  \eqref{Dpdt_singleDropCondense} reduces to the leading-order linear problem for $\pps\to P_L$ 
\begin{equation}
    \frac{d\pps}{dt} \sim -c_0 - c_1\left(\pps-\pinf\right)\qquad \implies\qquad
      \pps \sim \pinf - \frac{c_0}{c_1} + C e^{-c_1 t},
\lbl{eq:condensationLimit}
\end{equation}
where $c_0, c_1$ are positive constants depending on $\pinf, \pstar$.
The dot-dashed curve in Fig.~\ref{fig:SingleDropCondense}b confirms the exponential decay of $\pps$ as the contact lines of the condensing drop approach the edge of the domain.
\par
At a critical finite time, here $t_f \approx 1.8\times 10^8$, the entire domain is filled with liquid, and the dynamical model \eqref{Dpdt_singleDropCondense} no longer applies as the droplet has merged with its periodic images and starts approaching a spatially-uniform condensing film \cite{ji2018instability}.
Following a short transient at $t_f^+$  
the scaling law $\M(t) =Lh(t)= O(\sqrt{t-t_1})$ (see \eqref{flood}) for filmwise condensation applies with $\pps\sim \Pi(h(t))=O([t-t_1]^{-3/2})\to 0$, see Figs.~\ref{fig:SingleDropCondense}b,c.
Fig.~\ref{fig:SingleDropCondense}c also compares the evolution of $\M$ in rescaled time $\hat{t} = \beta t$ at different values of $\beta$. In all cases, we observe that $\M(\hat{t}\,)$ follows the predicted transition from $O(\hat{t}\,)$ to $O(\sqrt{\hat{t}}\,)$.
\par
Fig.~\ref{fig:SingleDropCondense}d shows results on droplet width vs.\ pressure from PDE simulations following that in Fig.~\ref{fig:SingleDropCondense}a at several different values of $\beta$. In all cases, the droplets retain the parabolic profile \eqref{eq:drop_parabolic}. For $\beta=O(\eps^3)$ and smaller, $w(\pps)$ coincides with the quasi-static relation \eqref{eq:drop_width}, which corresponds to a drop condensing with a constant effective contact angle,
$\Hd(w)=-\pps w = -A$ (c.f. \cite{wilson2023}). For larger $\beta$, the evolution of drops is different, as evidenced by the fact that they do not have a constant contact angle and relatedly their mass does not follow $M=O(\pps^{-2})$; we will not consider this range here.

\section{Two-droplet interactions}
\label{sec:twoDrop}
Next, we discuss the interactions between two droplets ($N = 2$) in a periodic domain. To separate the influences of pressure and inter-drop spacing on the droplet dynamics, we consider two simplified cases: (1) two droplets with identical pressures and unequal spacing, and (2) two droplets with identical spacing and unequal pressures.
Insights gained from the analysis of these reduced scenarios will be applied to systems with more droplets. 

\subsection{Equally-sized droplets with un-equal spacing: separation vs. collision}
\label{sec:twoDropEqualPressure}
\begin{figure}
    \centering
    \includegraphics[width=6cm]{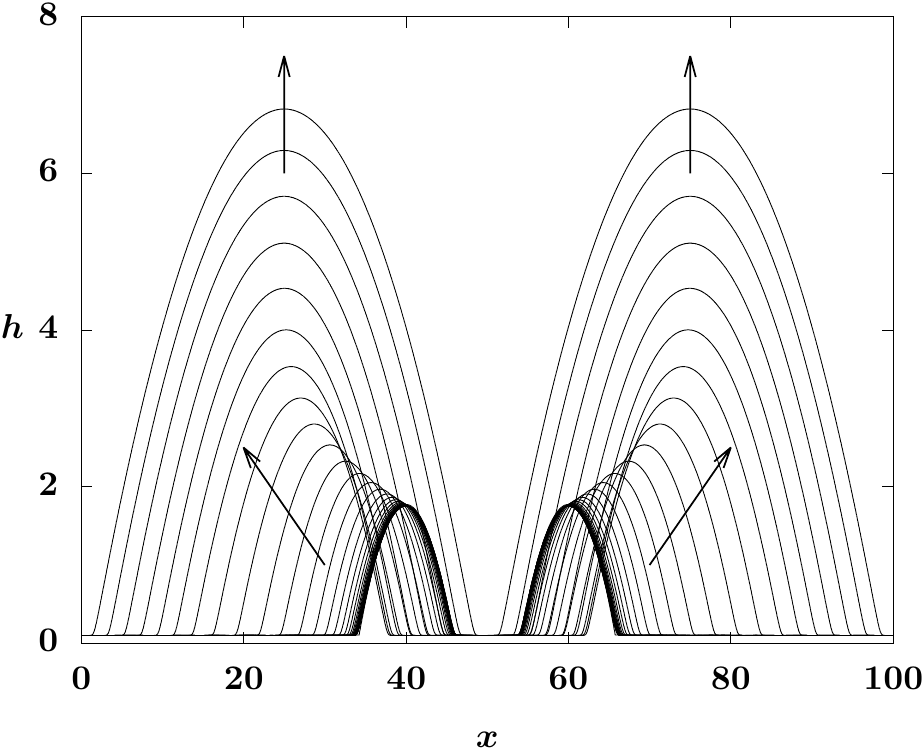}
    \includegraphics[width=6cm]{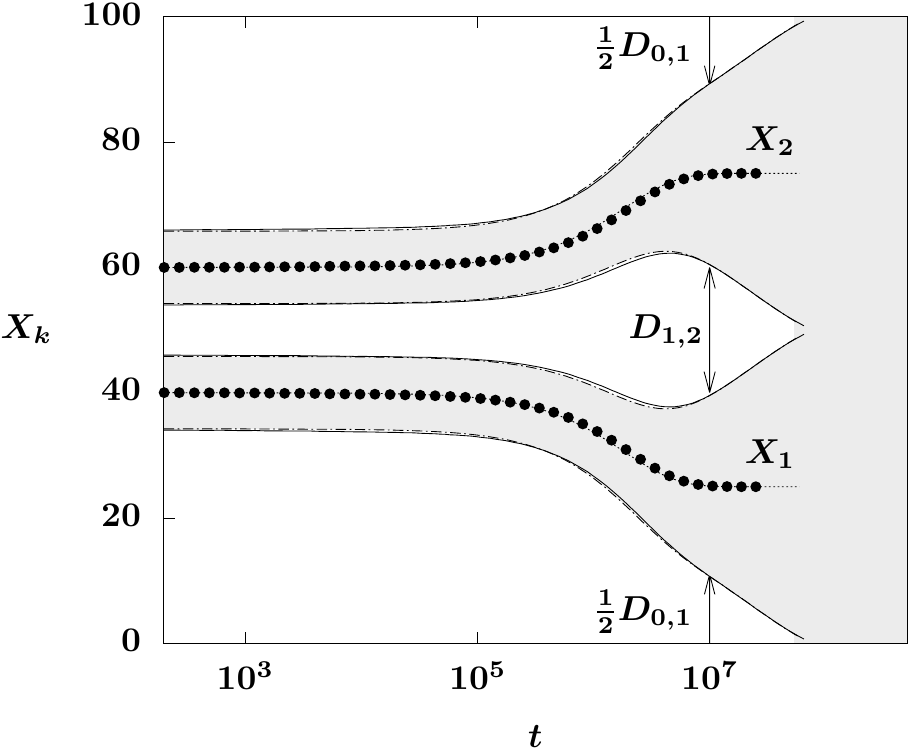}
    \caption{(Left) PDE simulation of \eqref{Mainpde} for the dynamics of two equally-sized droplets in a periodic domain.  At time $t=0$, two droplets with identical pressure $\pps=0.1$ are placed at $\X_1 = 40$ and $\X_2 = 60$. The droplets separate and become more equally spaced as weak condensation occurs. (Right) 
    The evolution of the droplet peak positions $\X_1(t)$ and $\X_2(t)$ (marked by dots) and their contact line positions  $\X_{1,2} \pm w(\pps)$ (solid curves) from the PDE simulation agree well with the predictions (dashed curves) from the dynamical model \eqref{eq:equalPressureSystem}. The domain size is $L=100$ and $\beta = 10^{-7}$. The shaded area shows the trajectory of two droplets, leading up to filmwise condensation after the droplets collide.
    }
    \label{fig:2Drop_Dx}
\end{figure}

Consider two droplets of identical pressure, $\pps_1 = \pps_2=\pps$, (and consequently equal masses \eqref{eq:drop_mass} and equal sizes, $w=A/\pps$) in a periodic domain $0<x<L$ (see Fig.~\ref{fig:2Drop_Dx}). Denote the distance between the right contact line of the first droplet and the left contact line of the second droplet as $D_{1,2} = (\X_2-w)-(\X_1+w)$. Since the domain is periodic, the distance between the right contact line of the second droplet and the left contact line of the first droplet is $D_{0,1} = D_{2,3} = L - (\X_2+w)+(\X_1-w)$.

Then the ODE system \eqref{eq:ODE_system} for the equally-sized droplets reduces to
\begin{subequations}
\lbl{eq:equalPressureSystem}
\begin{equation}
\label{eq:dpdt_equalDrop}
    {d\pps\over dt}=
    \beta C_P(\pps)(\pps-\pstar)\left[\frac{\mu}{2}(D_{1,2}+D_{0,1})+\frac{\Jnc(\pps)}{\pps-\pstar}\right],
\end{equation}
\begin{equation}
\frac{d \X_1}{dt} = -\frac{d \X_2}{dt} = -\frac{\beta\mu}{2}C_X(\pps)(\pps-\pstar)(D_{1,2}-D_{0,1}).
\end{equation}
\end{subequations}
Using $w=A/\pps$, we have
$$
    \frac{dD_{1,2}}{dt} = \frac{d\X_2}{dt}- \frac{d\X_1}{dt} + \frac{2A}{\pps^2}\frac{d\pps}{dt}
\qquad
\frac{dD_{0,1}}{dt} = -\frac{d\X_2}{dt}+ \frac{d\X_1}{dt} + \frac{2A}{\pps^2}\frac{d\pps}{dt}\, .
$$
Therefore, the difference between the inter-droplet spacing $\Delta D = D_{0,1}-D_{1,2}$ satisfies
\begin{equation}
 \frac{d}{dt}\Delta D = 2\beta\mu C_X(\pps)(\pps-\pstar)\Delta D.
\end{equation}
This analysis shows that if $\pps<\pstar$, then the magnitude of $\Delta D$ can be bounded by an exponential decay in time, making the droplets asymptotically approach equal spacing for long times. However, it can be shown that the $dD/dt$ evolution equations allow for merging of droplets at finite times, $D\to 0$ as in the single drop case from section~\ref{sec:singleDrop}. We note that the differences from the mass-preserving linear instability of periodic solutions described by \cite{LP} stems from the difference in the form of the non-conservative flux term in \eqref{Mainpde}. Here periodic droplet arrays are unstable to filmwise condensation.
\par
The PDE simulation in Fig.~\ref{fig:2Drop_Dx}(left) shows an example of the dynamics of two equally-sized droplets with unequal spacing. As the droplets condense, they move away from each other, becoming more equally spaced, approaching $\X_1\to L/4$ and $\X_2\to 3L/4$, and eventually merge into a single large droplet via droplet collision. 
The plot in Fig.~\ref{fig:2Drop_Dx}(right) traces the separation of the centers of the droplets at $\X_1$, $\X_2$, as well as the contact line motions of the droplets $\X_{1,2} \pm w$ in time. The results from the PDE simulation (dots for the peak positions and solid curves for the contact line positions) and from the reduced ODE system \eqref{eq:equalPressureSystem} (dashed curves) show good agreement. Based on the exponential decay of $\Delta D$ vs. the finite-time decay of the $D$'s, one pair of the contact lines will merge first ($D_{0,1}\to 0$ or $D_{1,2}\to 0$), slightly before the other (when the domain becomes completely filled).

\subsection{Equally-spaced droplets with different pressures: growth vs. collapse} 
\label{sec:twoDropEqualSpace}
Next, we consider two droplets of different pressures, $\pps_1$ and $\pps_2$, in a periodic domain $0\le x\le L$. Furthermore, we assume that the droplets are placed at the positions $\X_1 = L/4$ and $\X_2 = 3L/4$, respectively, so that the droplets are equally spaced, with $\lambda=\X_2-\X_1=L/2$, in the periodic domain. 
\par
\begin{figure}
    \centering
 \mbox{ (a) \includegraphics[width=3.6cm]{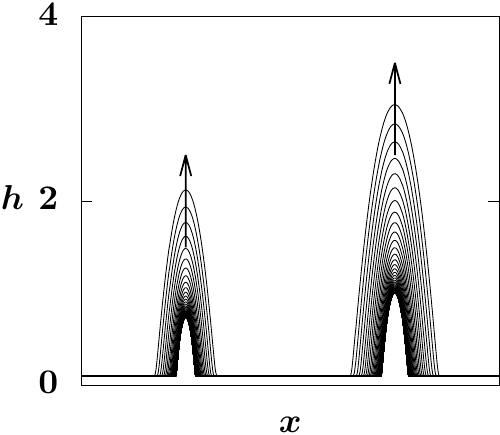} 
   (b) \includegraphics[width=3.6cm]{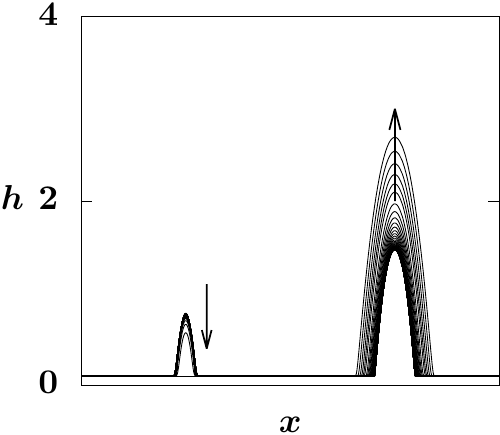} 
  (c)\includegraphics[width=3.6cm]{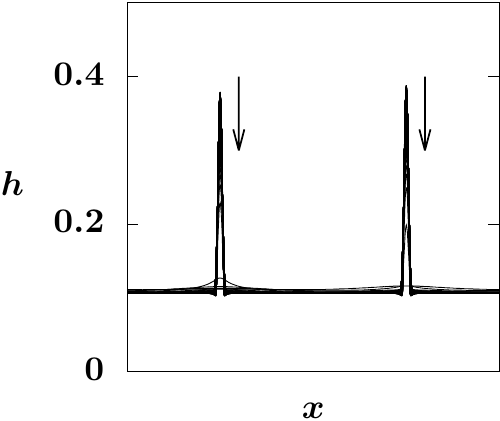} }
    \caption{Typical PDE simulations starting from two droplets of pressures $(\pps_1, \pps_2)$ placed at $\X_1=L/4$ and $\X_2 = 3L/4$ in a periodic domain $0\le x \le L$: (a) $(\pps_1, \pps_2) = (0.25, 0.18)$  corresponds to case (A) pairwise growing; (b) $(\pps_1, \pps_2) = (0.25, 0.12)$  corresponds to case (B) growing-shrinking; (c) $(\pps_1, \pps_2) = (0.62, 0.6)$  corresponds to case (C) pairwise shrinking.
   The system parameters are $L = 100$, $\beta=10^{-7}$.
    }
    \label{fig:twoDropPDEs}
\end{figure}

Figure~\ref{fig:twoDropPDEs} shows PDE simulations illustrating the three different generic modes of dynamics starting from two droplets:
\begin{enumerate}
    \item[(A)] \emph{Pairwise growing}: both droplets grow in time.
    \item[(B)] \emph{Growing -- shrinking}: one droplet shrinks and collapses in finite time, and the other droplet grows.
    \item[(C)]\emph{Pairwise shrinking}: both droplets shrink.
\end{enumerate}
For cases (A) and (C), non-conservative effects (condensation or evaporation respectively) must play a role since the overall mass is changing. For case (B), it may not be clear if the dynamics are dominated by conservative or non-conservative fluxes. While $\beta=10^{-7}$ used here may appear rather small, noting the separation distance $D=O(L/2)$ and recalling the discussion in Section~\ref{sec22}, it is close to (but smaller than) the critical $\beta_c=K\eps^3/D^2$, that limits when these two fluxes can be separated out as written in the model.

We first show that the droplets do not change positions.
Here the ODE system \eqref{dPk} with $N = 2$,
can be written as
\begin{eqnarray}
\frac{d\X_1}{dt} = -C_X(\pps_1)(D_{1,2}-D_{0,1})\left[{\frac{\Hm^3(\pps_2-\pps_1)}{D_{1,2}D_{0,1}}}+{\frac{\beta\mu}{6}}(2\pps_1+\pps_2-3\pstar )\right],\\
\nonumber
\frac{d\X_2}{dt} = -C_X(\pps_2)(D_{0,1}-D_{1,2})\left[{\frac{\Hm^3(\pps_1-\pps_2)}{D_{0,1}D_{1,2}}}+{\frac{\beta\mu}{6}}(\pps_1+2\pps_2-3\pstar )\right].
\end{eqnarray}
Due to the periodicity,  the distances between contact lines are identical,
\begin{equation}
D_{1,2}=D_{0,1} = \frac{L}{2} - \frac{A}{\pps_1}-\frac{A}{\pps_2} > 0.
\label{eq:equalDistance}
\end{equation}
Therefore, we have $d\X_1/dt=d\X_2/dt=0$,
which indicates that droplets do not move, and the dynamics of the system are only determined by the droplet pressures $(\pps_1, \pps_2)$. 
Again, using the ODE system \eqref{dPk} and \eqref{eq:equalDistance}, we have
\begin{eqnarray}
&&    \lbl{eq:equalSpacingPsystem}\\
\nonumber
    \frac{d\pps_1}{dt} &=&  C_P(\pps_1)\left[\frac{2\Hm^3 (\pps_1-\pps_2)}{D(\pps_1,\pps_2)}+{\frac{\beta\mu }{3}}D(\pps_1,\pps_2)(2\pps_1+\pps_2-3\pstar)+\beta \Jnc(\pps_1)\right],\\
\nonumber
    \frac{d\pps_2}{dt} &=&  C_P(\pps_2)\left[\frac{2\Hm^3 (\pps_2-\pps_1)}{D(\pps_2,\pps_1)}+{\frac{\beta\mu }{3}}D(\pps_2,\pps_1)(2\pps_2+\pps_1-3\pstar)+\beta \Jnc(\pps_2)\right],
\end{eqnarray}
where we write $D(\pps_1,\pps_2) = D_{1,2} = D_{0,1}$ as defined in \eqref{eq:equalDistance} to highlight the dependence of the contact line separation on the droplet pressures.

The system \eqref{eq:equalSpacingPsystem} is an autonomous phase plane system for $\pps_1, \pps_2$ with an equilibrium point at $\pps_1=\pps_2=\pstar$ that is an unstable node. Formally, the system has another equilibrium point on the line $\pps_2=\pps_1$ but this point is inaccessible (a ``virtual'' point  \cite{budd}) since it lies in the unphysical region with $D(\pps_1,\pps_2)<0$. Still, it has a strong influence on the structure of the physically relevant solutions for $D>0$.
On the line $\pps_2=\pps_1$, the system reduces to the one-drop dynamics given by \eqref{Dpdt_singleDropCondense} with the domain being mapped to $L\to L/2$ and  $\pinf=4A/L$. 

\begin{figure}
\centering
\mbox{
(a)\includegraphics[width=0.43\textwidth]{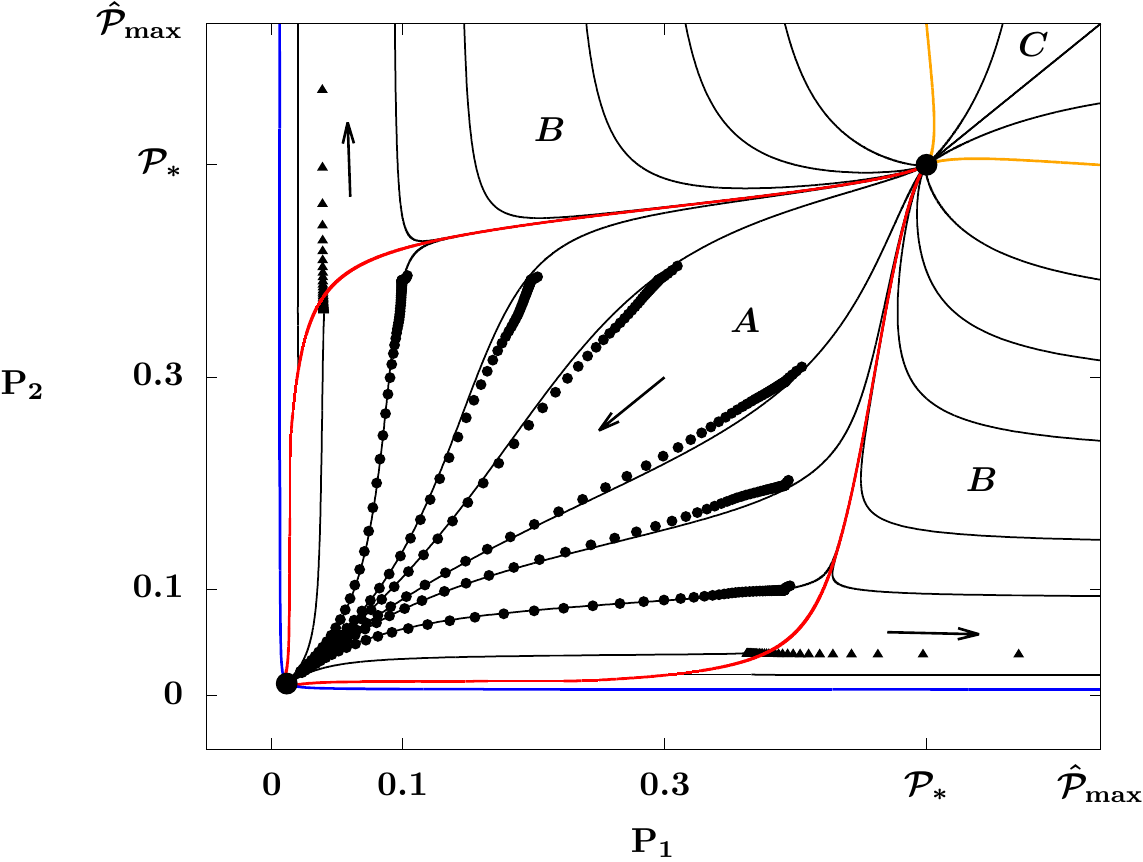}
(b)\includegraphics[width=0.43\textwidth]{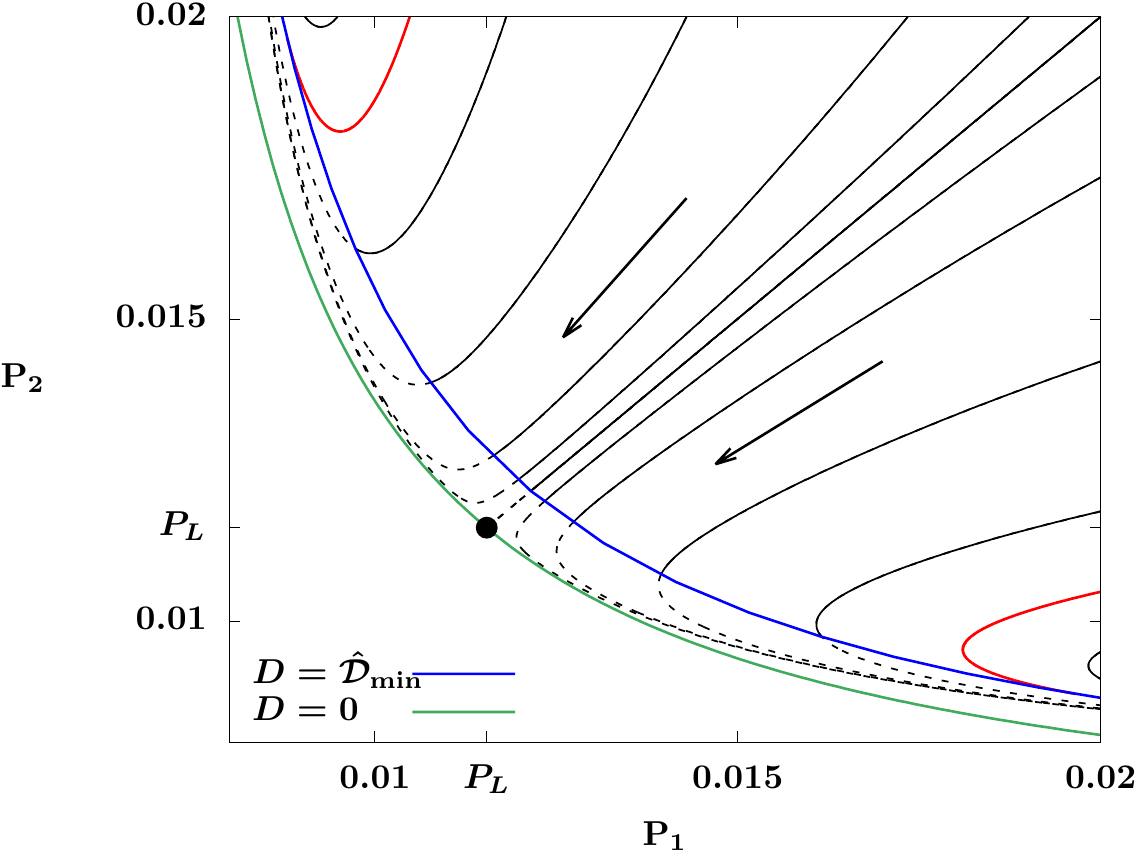}
}
\mbox{} \\
\mbox{
(c)\includegraphics[width=0.43\textwidth]{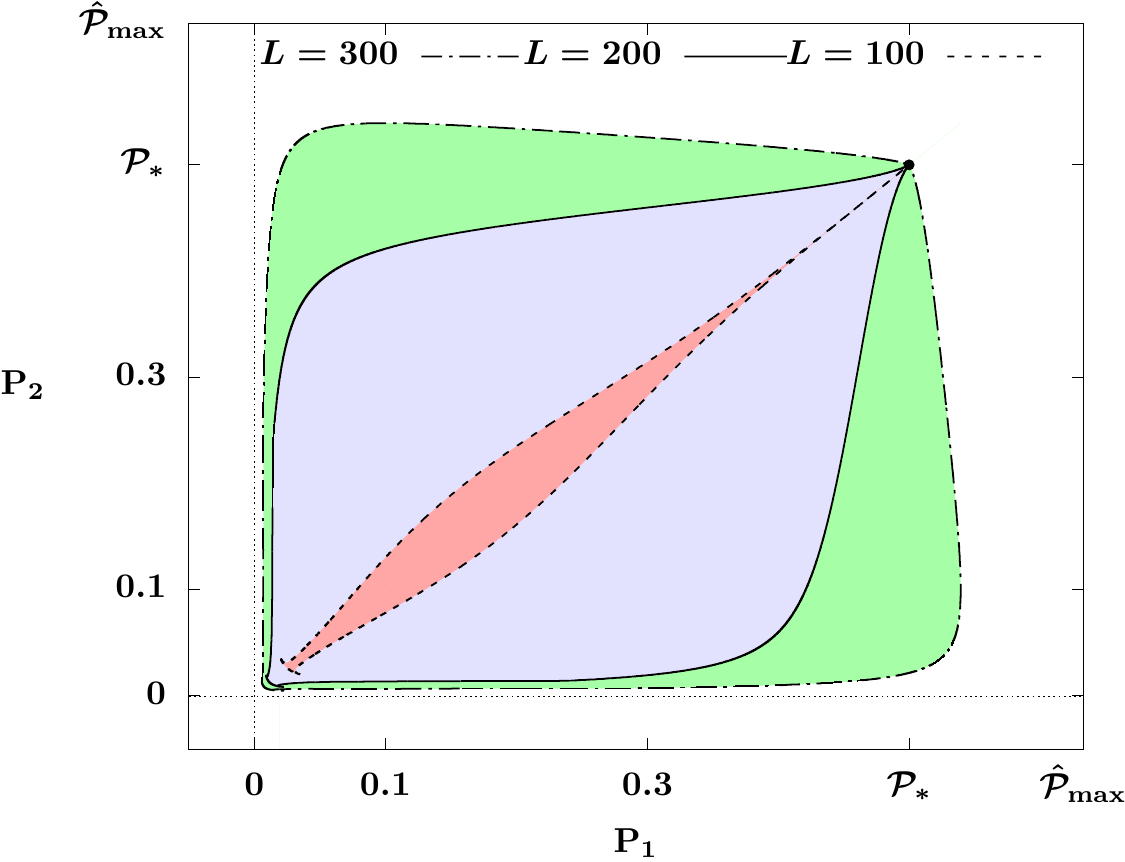}
(d)\includegraphics[width=0.41\textwidth]{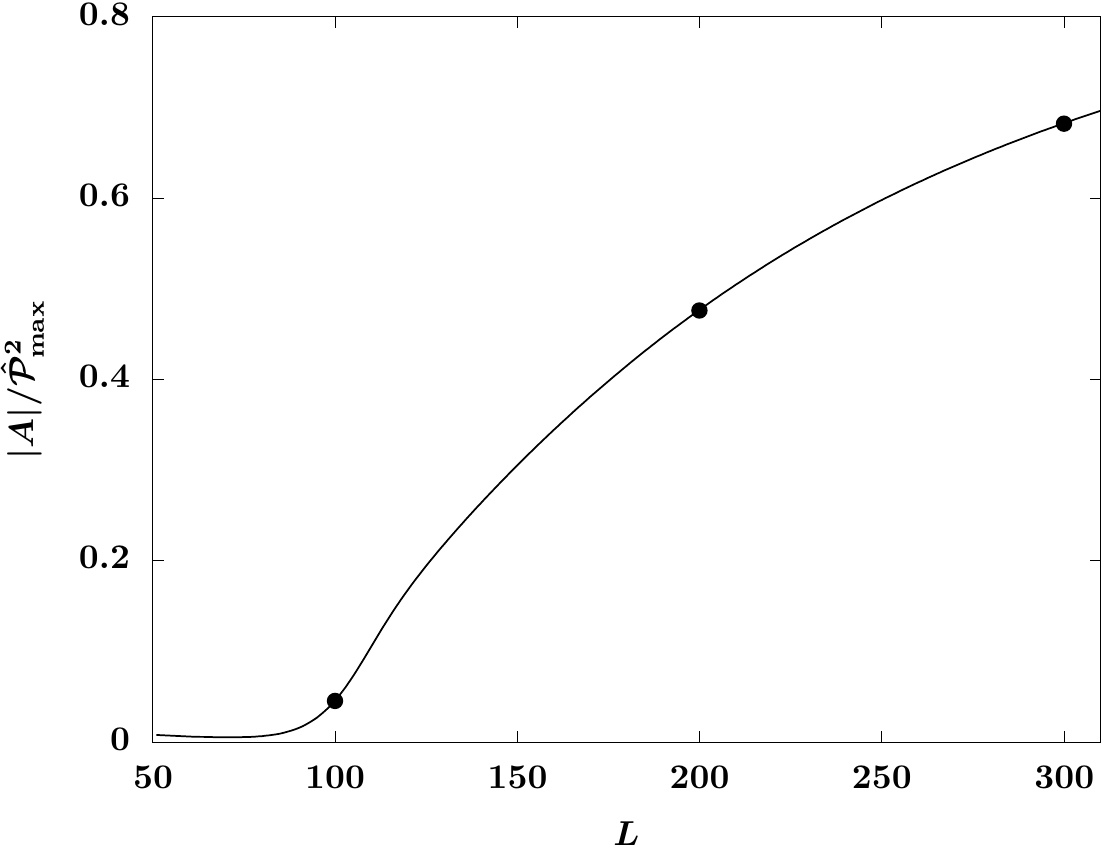}
}
\caption{
(a) ODE solution trajectories (black) compared against PDE solutions (dots) for the equally-spaced two droplet case with $L = 200.$ The separatrix between regions A/B (red), the separatrix between regions B/C (orange), and the collision criteria (blue) \eqref{eq:collision_criteria} identify three droplet dynamic regions.
(b) A close-up view of Fig.~(a), where the black dashed curves represent the unphysical trajectories that pass the $D_{1,2}=\Dmin$ collision criteria (plotted in blue). The red dotted curve represents $D_{1,2}=0$.
(c) Region (A) for $L=100, 200, 300$ (corresponding to three dots in Fig.~(d)). (d) Area of region (A) (pairwise growing) as a function of the domain size $L$. The other parameters are $\beta=10^{-7}, \Dmin=5, \pmaxh=0.63$.
}
\label{fig:twoDropTrajectories}
\end{figure}

The curve $D(\pps_1, \pps_2)=0$ makes the first terms on the right in the system \eqref{eq:equalSpacingPsystem} singular, hence this is not the most convenient form for standard analysis. Consequently, to remove this singularity, following  \cite[Sect 11.4]{murray}, we define a new time variable by the
relation ${ds/dt}= 1/ D(\pps_1, \pps_2)$
then \eqref{eq:equalSpacingPsystem} can be re-written as 
{\small
\begin{eqnarray}
 \lbl{eq:equalSpacingPsystem2}
 &&\\
 \nonumber
    \frac{d\pps_1}{ds} &=&  C_P(\pps_1)\left[{2\Hm^3 (\pps_1-\pps_2)}+D(\pps_1,\pps_2)\left\{{\frac{\beta\mu }{3}}D(\pps_1,\pps_2)(2\pps_1+\pps_2-3\pstar)+\beta \Jnc(\pps_1)\right\}\right],
        \\
    \nonumber
    \frac{d\pps_2}{ds} &=&  C_P(\pps_2)\left[{2\Hm^3 (\pps_2-\pps_1)}+D(\pps_2,\pps_1)\left\{{\frac{\beta\mu }{3}}D(\pps_2,\pps_1)(2\pps_2+\pps_1-3\pstar)+\beta \Jnc(\pps_2) \right\}\right].
\end{eqnarray}
}
In this form, the system has a hyperbolic saddle point at $\pps_2=\pps_1=\pinf$. The stable manifold of 
this point is the line $\pps_2=\pps_1$, and the unstable manifold is the hyperbola given by $D(\pps_1,\pps_2)=0$ (see Fig.~\ref{fig:twoDropTrajectories}b).

Figure~\ref{fig:twoDropTrajectories}a shows a typical $(\pps_1, \pps_2)$ phase plane of the system \eqref{eq:equalSpacingPsystem2} with $L = 200$ and $\beta=10^{-7}$. Three regions appear in the phase plane, where region (A) describes pairwise droplet growth, region (B) describes growing-shrinking dynamics, and region (C) characterizes pairwise shrinking.

\par
First, we discuss the separatrix between region (A) and region (B).
The solid dots in Fig.~\ref{fig:twoDropTrajectories}a correspond to PDE simulations starting from initial droplet pressures in region (A), showing good agreement with the pressure trajectories (black solid curves) from the dynamical system \eqref{eq:equalSpacingPsystem2} leading to pairwise condensation. When the droplet pressures are close to the A/B separatrix (plotted in red), PDE solutions can deviate from the ODE trajectories; one such example is presented by the solid triangles in Fig.~\ref{fig:twoDropTrajectories}a where the corresponding ODE trajectory indicates that both droplets should grow, while the PDE simulation yields growing-shrinking dynamics.
\par
For short to moderate times, we see in Fig.~\ref{fig:twoDropTrajectories}a that all of the trajectories in region (A) appear to have both pressures decreasing, but a close-up view  near the hyperbolic saddle point at $\pps_2=\pps_1=\pinf$ in Fig.~\ref{fig:twoDropTrajectories}b shows that only the stable manifold trajectory $\pps=\pps_1=\pps_2$ hits the $D=0$ collision curve. For sufficiently long times, all other trajectories bend away (parallel to the unstable manifold) to become growing-shrinking dynamics. Hence the regularized collision condition  $D(\pps_2,\pps_1) = \Dmin$ in \eqref{eq:collision_criteria}  plays a crucial role in determining the extent of region (A). 
\par
Trajectories that intersect the curve $D = \Dmin$ transversely terminate there, corresponding to droplet collisions (continuations in time have $D<\Dmin$ and are unphysical (shown by dashed curves)). 
The outermost trajectory in region (A) is the separatrix between A/B regions (red in Fig.~\ref{fig:twoDropTrajectories}b), identified as the trajectory that is tangent to the curve $D=\Dmin$ at a finite point. Further trajectories do not intersect $D=\Dmin$ and are part of region (B).
While the separatrix curves of the ODE system may not robustly separate the phase space for complex dynamics involving multiple droplets or for the PDE system, the pairwise droplet dynamics provide valuable insights into the general system behavior.

The dynamic regions are also dependent on the domain size $L$. Fig.~\ref{fig:twoDropTrajectories}c shows a comparison of region (A) (pairwise growth) for a varying domain size $L = 100,200,300$ (with other parameters fixed).
It indicates that region (A) enlarges when the domain size $L$ increases.
We can use the area of region (A) relative to $\pmaxh^2$ (the area of the phase plane for all droplet pair initial conditions) as a heuristic estimate for the likelihood of occurrence of pairwise-growth dynamics from generic initial conditions (see Fig.~\ref{fig:twoDropTrajectories}d). This indicates that pairwise condensations becomes the dominant behavior for more widely-separated droplets.

The separatrix between regions B/C depends on the collapse condition \eqref{eq:collapseCondition}. Consider a coupled droplet system where one of the droplets collapses in finite time at $t=t_c$, with $\pps_1 = \pmaxh$. Then the coupled droplet system reduces to the single-droplet dynamics following \eqref{Dpdt_singleDropCondense}. The remaining droplet will grow if its pressure $\pps_2 < \pstar$ at $t=t_c$, leading to growing-shrinking dynamics described by region (B). In contrast, if $\pps_2 > \pstar$ at $t=t_c$, then the remaining droplet  will also collapse in finite time, leading to the pairwise shrinking dynamics in region (C). Therefore, we identify the separatrix between B/C regions by the trajectories passing through the points $(\pps_1,\pps_2)=(\pmaxh, \pstar)$ or $(\pps_1,\pps_2)=(\pstar, \pmaxh)$ (orange curves in Fig.~\ref{fig:twoDropTrajectories}a).

\par
Using the definition of $C_P(\pps)$, \eqref{Eq:dpdt_scaled}, we can also  estimate the evolution for the difference in droplet masses, $\Delta {M} = {M}(\pps_2) - {M}(\pps_1)$.
For droplet pairs in the region (B) where one drop condenses and the other one collapses, it is clear that the mass difference increases in time.
For droplets in the region (A) that undergo pairwise condensation, it can be shown the larger droplet gains mass more quickly than the smaller drop, leading to an increasing absolute difference in mass. However, 
while the absolute differences in mass, $\Delta {M}$ grows, the relative difference $(\Delta {M})/{M}$ decreases.
\par
For droplets in the region (C) that collapse via pairwise evaporation, the smaller droplet loses mass more rapidly than the larger droplet, again leading to an increasing difference in mass.
It is also worth noting that the monotonicity of the mass difference does not imply that the pressure difference between droplet pairs also evolves monotonely.

\section{Many-droplet systems}
\lbl{sec:many_drop}
\begin{figure}  
    \centering
    \mbox{
    (a)\includegraphics[width=0.43\textwidth]{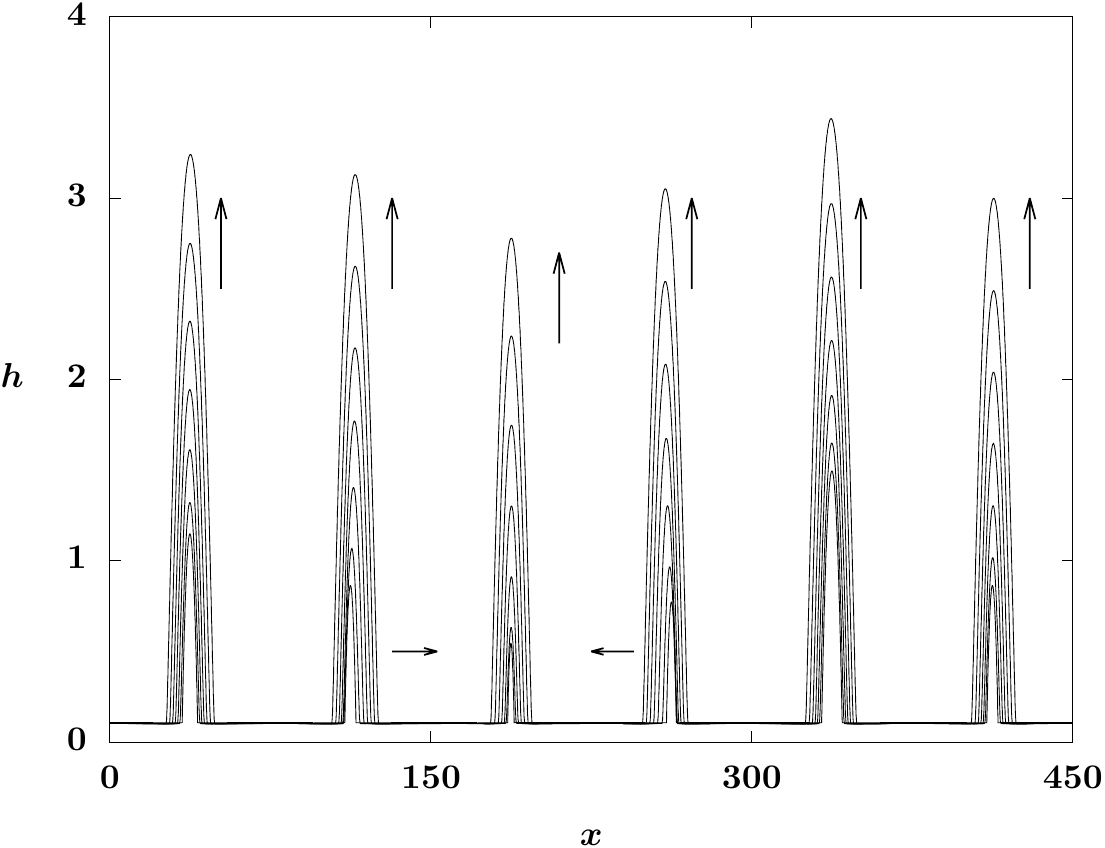}
   (b) \includegraphics[width=0.43\textwidth]{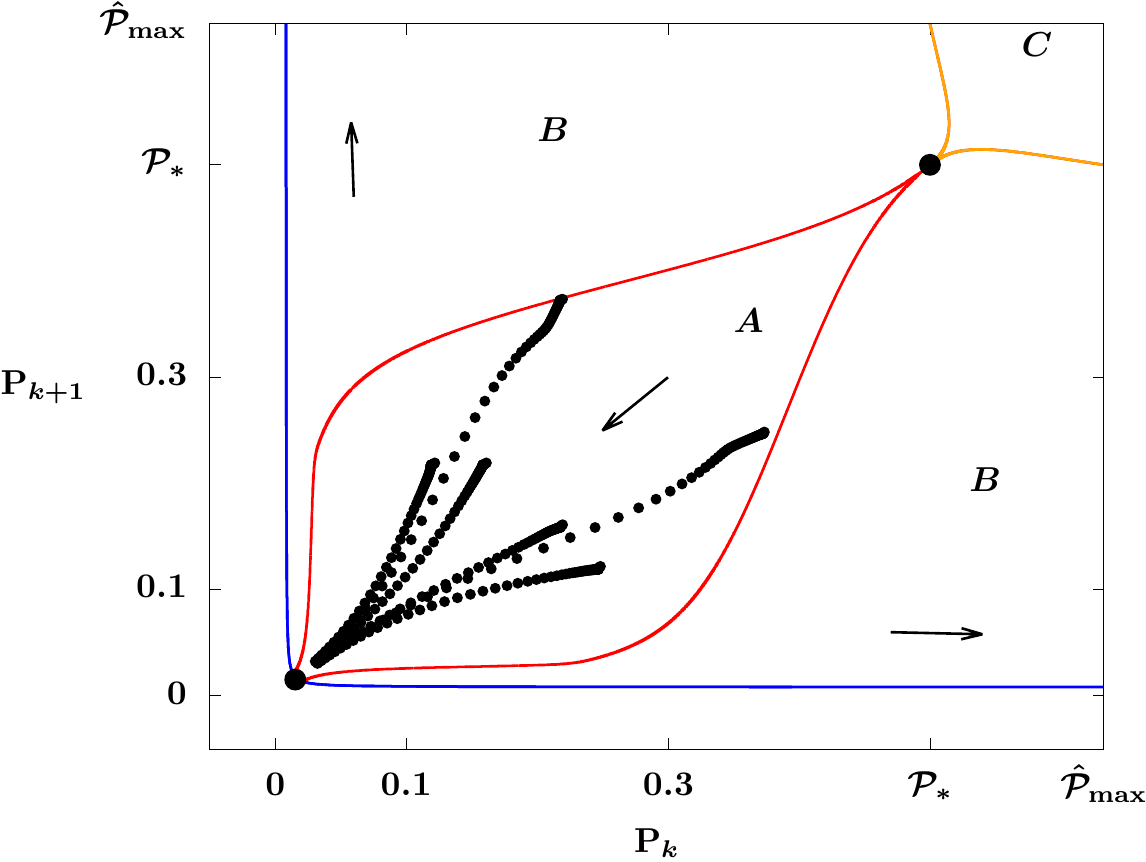}
    }
    \mbox{} \\
    \mbox{
(c)    \includegraphics[width=0.43\textwidth]{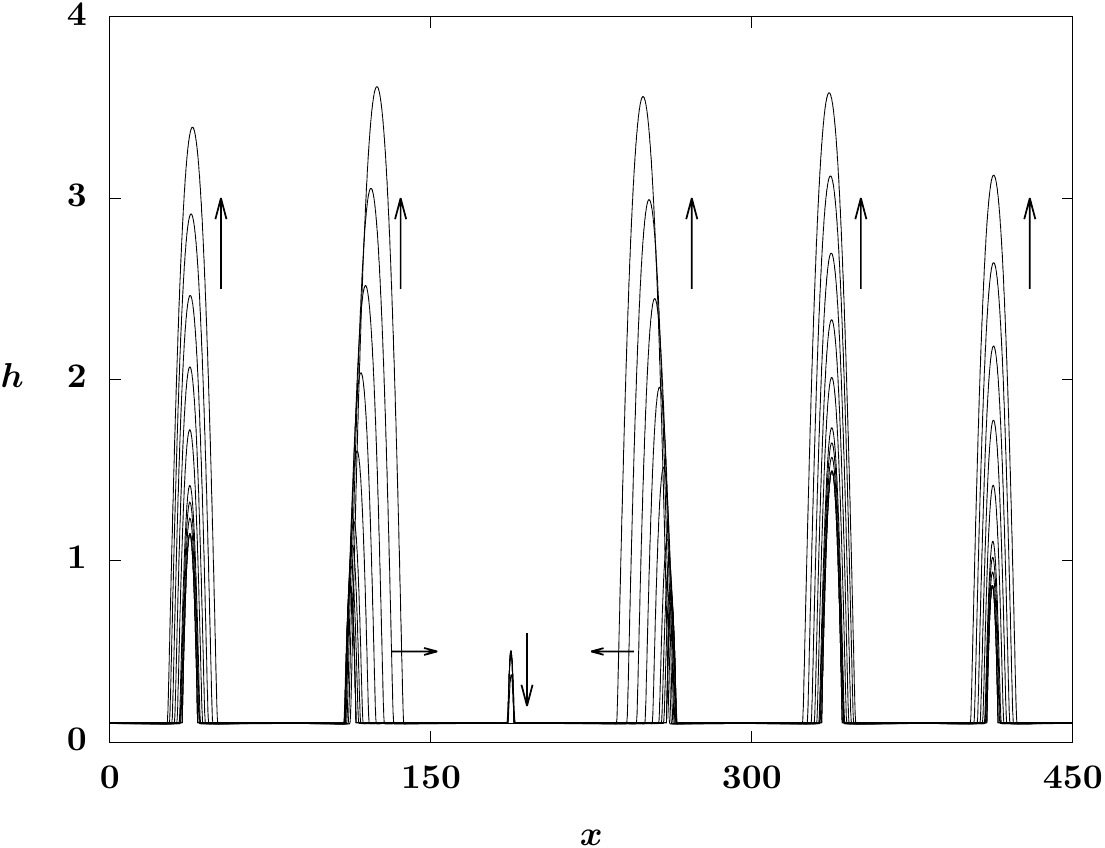}
   (d) \includegraphics[width=0.43\textwidth]{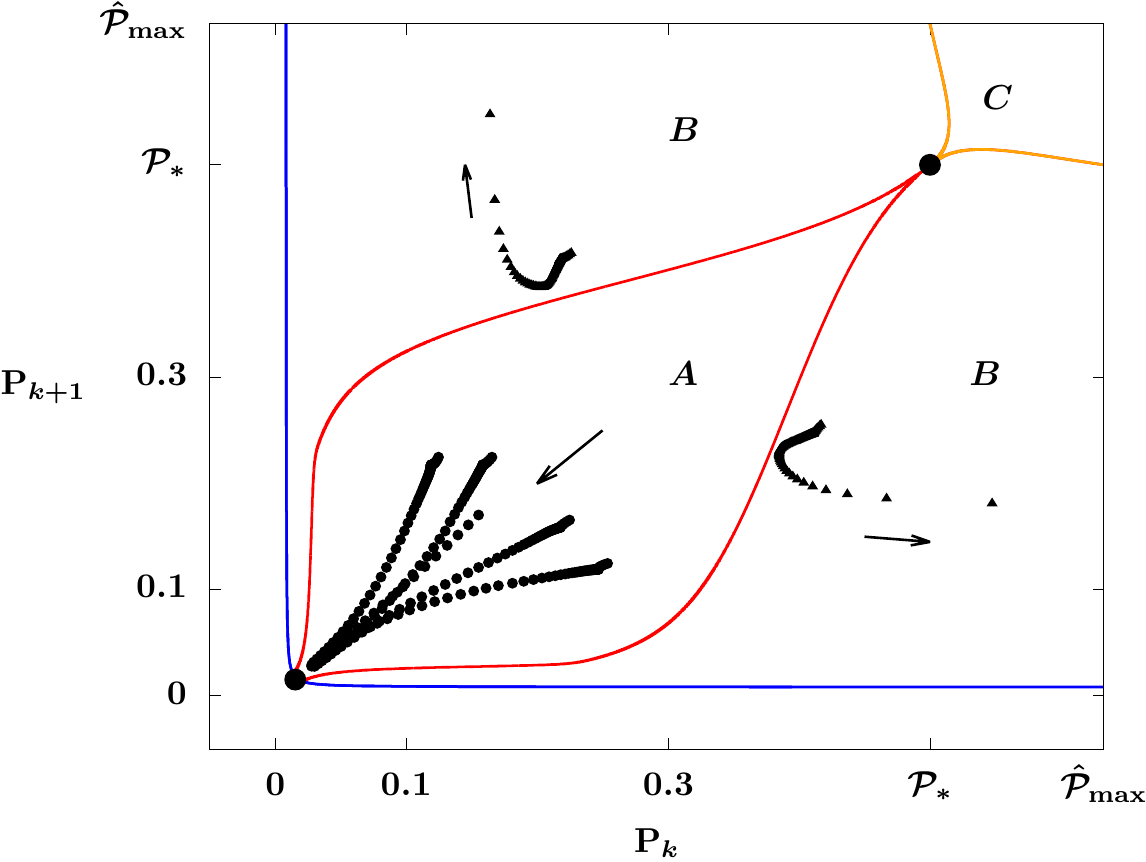}
   }
    \caption{PDE simulations and pairwise droplet pressure trajectories $(\pps_k(t), \pps_{k+1}(t))$ with $6$ droplets initially placed with equal spacing in a periodic domain $0\le x \le 450$. In the top panel (a,b), all the droplets grow in time, with pressure trajectories in the pairwise growth region (A). In the bottom panel (c,d), one droplet collapses and the other five droplets grow in time, yielding two pressure trajectories in the growing-shrinking region (B) and the other trajectories in region (A). The initial peak-to-peak spacing $\ppspacing=75$. The other parameters are identical to those used in Fig.~\ref{fig:twoDropTrajectories}, $\beta=10^{-7}$, $\Dmin=5$, $\pmaxh=0.63$. }
    \label{fig:DropPhasePlane_6Drops}
\end{figure}
For periodic systems with more than two droplets, lack of symmetry between neighboring drops can lead to more complicated scenarios, but the results from $N=2$ can still provide good guides for nearest-neighbor interactions.
Fig.~\ref{fig:DropPhasePlane_6Drops} presents the dynamics of a system in a periodic domain $0 \le x \le 450$ with $6$ equally-spaced droplets at initial positions $\X_k = (k-\textstyle\frac{1}{2})\ppspacing$, where $k = 1, \cdots, 6$, the peak-to-peak spacing $\ppspacing  = 75$, and pressures $\pps_k=( 0.16, 0.22, 0.38, 0.25, 0.12, 0.22)$.

PDE simulation of \eqref{Mainpde} shows that this initial configuration leads to growth of all droplets (see Fig.~\ref{fig:DropPhasePlane_6Drops}a). 
As the droplets grow in time, spatial motions also occur, hence the drops are no longer equally spaced with a constant peak-to-peak spacing. However, since the droplets are still widely separated, we approximate this many-droplet system by $6$ pairwise 
interactions between two neighboring droplets with positions $(\X_k(t), \X_{k+1}(t))$ and pressures $(\pps_k(t), \pps_{k+1}(t))$, for $k=1,\cdots,6$. 
Due to the periodic boundary conditions, the right-most drop with $k=6$ also forms a droplet pair with the left-most drop with $k=1$. Here, the initial droplet peak-to-peak spacing $\ppspacing$ is mapped to the domain size $L$ of a two-droplet dynamical system by $L = 2\ppspacing$.
Fig.~\ref{fig:DropPhasePlane_6Drops}b plots the trajectories of the paired pressures in the $(\pps_k(t), \pps_{k+1}(t))$ phase plane discussed in Fig.~\ref{fig:twoDropTrajectories}. This plot shows that all the droplet pairs fall into the pairwise growth region $(A)$.

Next, we slightly perturb the initial condition and change the pressure of the third droplet from $\pps_3 = 0.38$ to $\pps_3 = 0.42$, while keeping all the other positions and pressures unchanged. The PDE simulation result starting from the perturbed initial condition is shown in Fig.~\ref{fig:DropPhasePlane_6Drops}c. In this case, instead of growing, the third droplet  ($k=3$) collapses in finite time, and its two neighboring droplets move towards it at a faster rate. The droplets further away from the collapsed droplet do not show significant changes in dynamics compared to their counterparts in Fig.~\ref{fig:DropPhasePlane_6Drops}a. The trajectories of paired pressures for neighboring drops are also presented in the $(\pps_k, \pps_{k+1})$ phase plane in Fig.~\ref{fig:DropPhasePlane_6Drops}d, showing that two droplet pairs undergo growth-collapse (region (B)). We note that the collapsed droplet shows up in two trajectories as it forms droplet pairs with its left and right neighbors ($k=2,3$ and $3,4$).

This illustrates that for an array of well-separated droplets with nearly constant spacing and insignificant droplet drifts, the pairwise pressure phase plane from Section~\ref{sec:twoDropEqualSpace} for $(\pps_k, \pps_{k+1})$ provides a reasonable estimate for the droplet dynamics. When a droplet in the system collapses, the comparison between Fig.~\ref{fig:DropPhasePlane_6Drops}a and Fig.~\ref{fig:DropPhasePlane_6Drops}c also indicates that the influence of the collapsing drop on the rest of the system is weak. 

For many-droplet systems with significant spatial motions, further analysis of the coupled droplet dynamic model \eqref{eq:ODE_system}--\eqref{postcollide} is needed to describe the full dynamics. 
However, for case of weak condensation with large droplets, we have observed that this model provides good predictions for the dynamics.

For the next section, we will turn to the case of a system with a very large number of drops and consider the full coarsening dynamics going from $N\to N-1\to N-2 \to \cdots\to 1$ and filmwise condensation after that.

\section{Scaling regimes for coarsening  with weak condensation}
\label{sec:scaling}

\begin{figure}
    \centering
        \includegraphics[width=6.3cm]{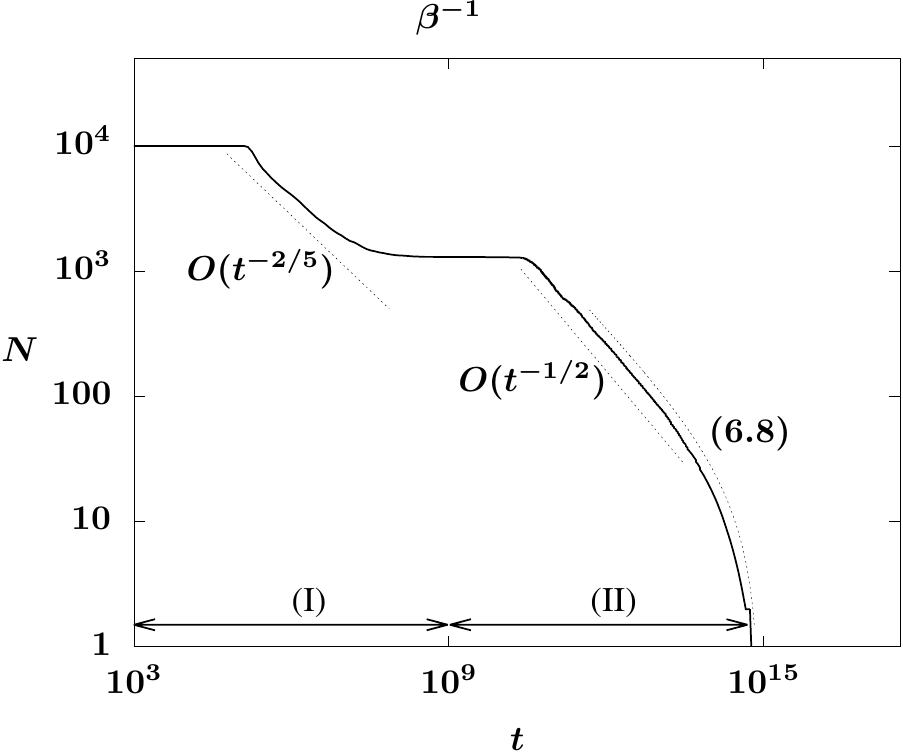}\quad
        \includegraphics[width=6.3cm]{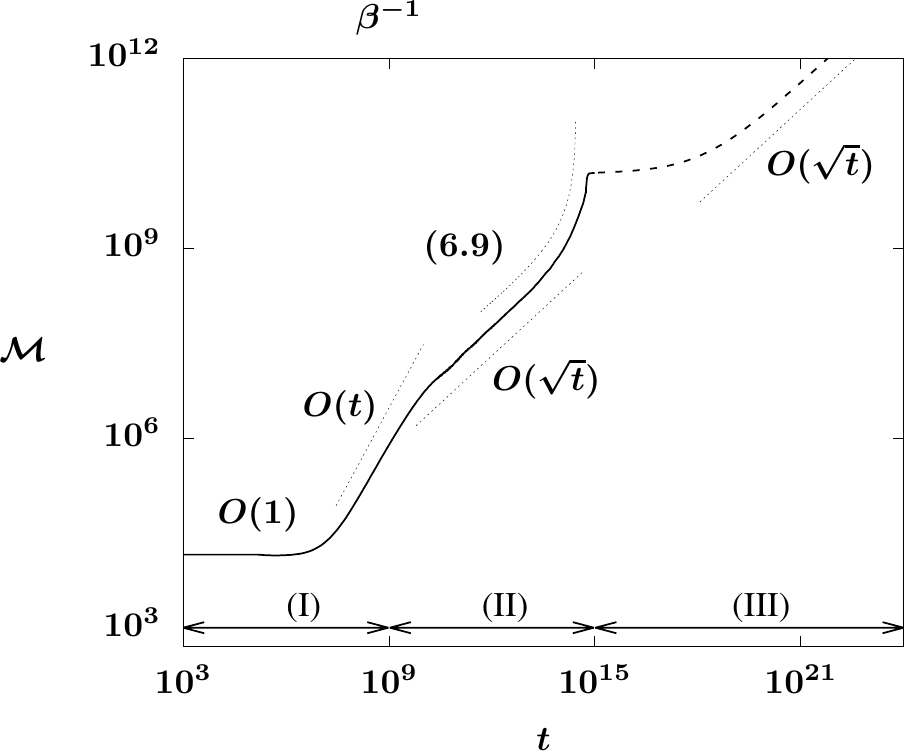}
    \caption{ Simulation of the augmented dynamical system \eqref{eq:ODE_system} starting from $N = 10^4$ droplets, showing a typical coarsening dynamics with three stages: (I) quasi-mass-conserving regime; (II) drop-wise condensation regime; (III) filmwise condensation regime.
    (Left) The coarsening rate shows a transition from $N = O(t^{-2/5})$ to $N = O(t^{-1/2})$, followed by logarithmic behaviors governed by \eqref{eq:N_log} as $N\searrow  1$. 
    (Right) The total mass of droplets in the system $\mathcal{M}(t)$ increases in time shows a transition from $\mathcal{M}= O(t)$ to $\mathcal{M}= O(\sqrt{t})$, followed by a logarithmic law \eqref{eq:M_log}.
    The dashed curve corresponds to filmwise condensation with $\mathcal{M}= O(\sqrt{t})$. 
    The initial droplet configuration has $\lambda = 40$, $\pavg=0.3$, and the other parameters are $\beta=10^{-9}$, $\Dmin=1$, $\pmaxh = 1$. 
    }
    \label{fig:N_M_t}
\end{figure}

To understand the dynamics in macro-scale physical problems where there may be a very large number of drops, $N\gg 1$, we now seek to heuristically obtain predictions for the new regimes of long-time asymptotic behaviors in the dynamics with weak condensation.
\par
 As an example we consider a typical simulation of the hybrid dynamical system \eqref{eq:ODE_system}--\eqref{postcollide} starting from an array of $N = 10^4$ droplets. The droplets are chosen to be initially equally-spaced at $t = 0$ with peak-to-peak spacing $\lambda = 40$ and pressures that are independent random variables, all following a uniform distribution with mean $\pavg=0.3$ and $\delta=0.02$,  $\pps_k(0)= \mathcal{U}(\pavg-\delta, \pavg+\delta)$.
The drops shown Fig.~\ref{fig:schematic}(left) can represent a typical set of three adjacent drops from this large array. 
Figure~\ref{fig:N_M_t} illustrates the long-time coarsening dynamics of this system in terms of both the number of droplets $N(t)$ and the total mass of droplets $\mathcal{M}(t)$  as functions of time.  We observe in Fig.~\ref{fig:N_M_t} (left) that as droplet collapse and collision events occur, the number of droplets decreases in time, with a transition from an early-stage scaling $N = O(t^{-2/5})$ (as seen previously in \cite{glasner2003coarsening,glasner2005collision}) to a later-stage scaling $N = O(t^{-1/2})$, followed by a logarithmic scaling as $N \searrow 1$. Correspondingly, the total mass of droplets $\mathcal{M}$ stays nearly constant for $t < 10^6$ ($t\ll O(1/\beta)$) and starts increasing as droplet condensation dominates the system in the later stage (see Fig.~\ref{fig:N_M_t}(right)). 
For the transitional regime between the two stages, $10^7\lesssim t \lesssim 10^{10}$, we observe that the number of droplets remains nearly constant, whereas the total droplet mass exhibits significant growth following $\mathcal{M} = O(t)$.
The value $\beta=10^{-9}$ used here was chosen to help more clearly separate the earlier conservative coarsening from the longer-time condensation-dominated dynamics.
\par
\begin{figure}\centering
\includegraphics[height=4.5cm]{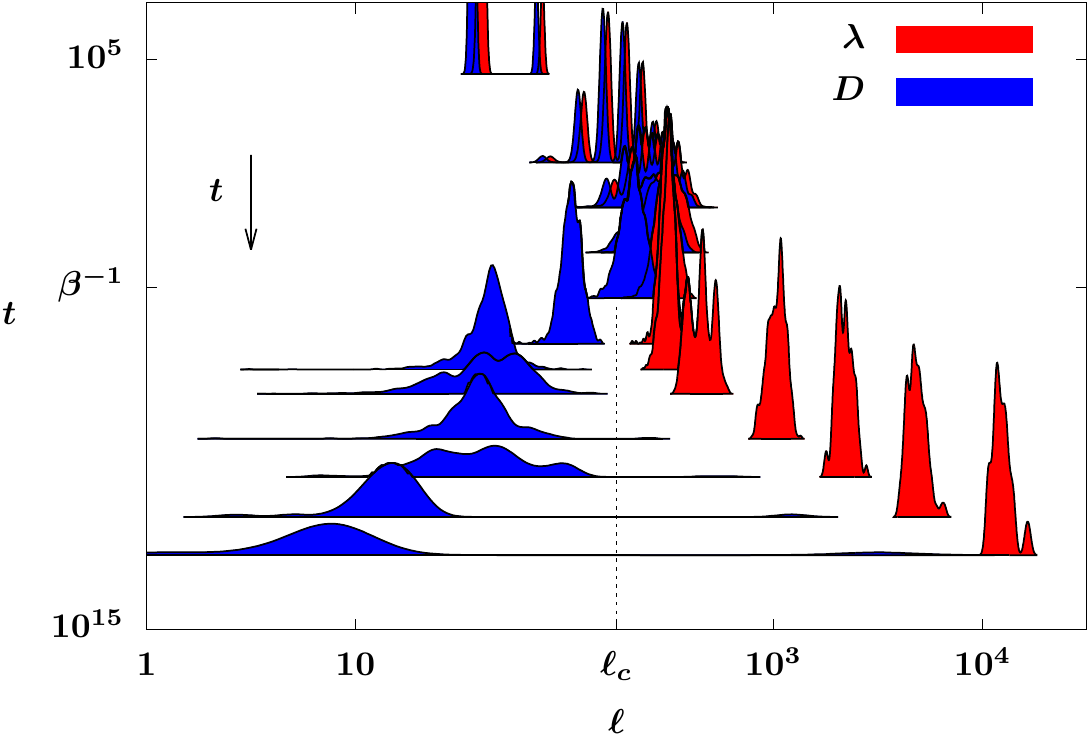}\qquad
\includegraphics[height=4.5cm]{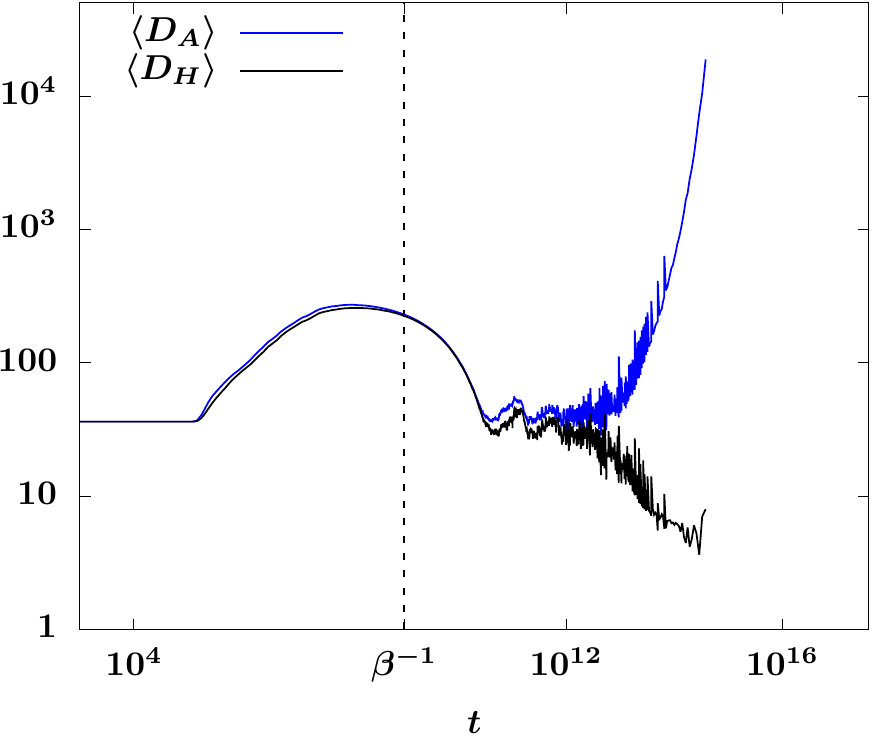}
\caption{(Left) Distribution plots of separation distances $\ell$ for droplets in time, in terms of 
peak-to-peak distances $\lambda_k=X_{k+1}-X_k$ and contact-line distances $D_k$, from the simulation in Fig.~\ref{fig:N_M_t}. (Right) Time-evolution of the arithmetic average $\bra D_A\ket$ and harmonic average $\bra D_H\ket$ of contact-line distances $D$.}
\label{fig:D_lamb_dist}
\end{figure}

In this section, we will focus on the discussion of these scaling laws for the coarsening dynamics.
Figure \ref{fig:D_lamb_dist} (left) provides another representation of the droplet system configuration as it evolves over time during coarsening, resulting in significant changes in the distributions of peak-to-peak spacing $\lambda$ and contact line distances $D$.
For $t < 10^8$, the average peak-to-peak spacing $\left<\lambda\right>$ and the average contact line distance $\left<D\right>$ are close to each other. This suggests that during the early stage, the majority of droplets in the system are relatively small in size, with widths that are small compared to the distance between droplets, hence collisions can be expected to be rare. As coarsening proceeds, $\left<\lambda\right>$ deviates from $\left<D\right>$ and increases in time.
We have shown in section~\ref{sec:twoDropEqualSpace} the droplet dynamics strongly depend on the peak-to-peak spacing $\lambda$, with pairwise droplet condensation (previously called case (A)) becoming the expected mode of dynamics when $\lambda$ is large (see Figs.~\ref{fig:twoDropTrajectories}c,d).  In Fig.~\ref{fig:D_lamb_dist}(left) we have indicated a line $\lambda=\ell_c\approx 225$ that was observed to be correspond to a strong transition in the coarsening to yield only droplet collision events for all later times (with $\bra \lambda \ket$ also continuing to monotone increase). This was observed to occur after the timescale expected for condensation effects to become dominant, $t=O(\beta^{-1})$.
\par
For what follows, it is useful to more carefully examine the distribution of 
separation distances between droplets.
Fig.~\ref{fig:D_lamb_dist}~(right) displays both the arithmetic and harmonic averages 
of the contact line distances $D_k$,
$$\bra D_A\ket = {1\over N}\sum_{k=1}^N D_k,\qquad \bra D_H\ket = N\bigg/\sum_{k=1}^N D_k^{-1},$$
having dramatic differences during the later stage. 
For $t > 10^{12}$,  $\bra D_A \ket$ increases while $\bra D_H\ket$ decreases, which is consistent with the distribution plots of $D_k$ in Fig.~\ref{fig:D_lamb_dist}~(left), showing the coexistence of a small number of very large $D_k$ values with the majority of distances being small.
\par
Motivated by the observations from Figure \ref{fig:N_M_t}, we now present heuristic arguments describing the three distinct stages of dynamics:
\par
\underline{\emph{Stage I: Quasi-conservative dynamics.}}
 In the early stage for $t \ll O(\beta^{-1})$, the system behaves like the mass-conserving coarsening case discussed in \cite{glasner2003coarsening,glasner2005collision,glasner2009ostwald}, as condensation has not yet had enough time to contribute significantly.
The average peak-to-peak spacing between droplets $\left<\ppspacing\right> \sim L/N$ is small, and the average size of  droplets is also small. In this case, Fig.~\ref{fig:twoDropTrajectories}c suggests that growing-shrinking dynamics occurs for a wider range of droplet pairs.
Therefore, the coarsening is dominated by droplet collapse due to mass exchange modified by weak condensation effects. The number of droplets $N$ follows the mass-conserving scaling, $N\bra M \ket=\mathcal{M}$, 
and the average droplet mass $\bra M \ket\propto \bra 1/\pps^2\ket$ from \eqref{eq:drop_mass}. 
Using Jensen's inequality, we have $\bra\pps\ket^2 \ge \bra 1/\pps^2\ket^{-1}$, 
which yields an upper bound for $N$ in terms of the average pressure, 
$N=O(\left<\pps\right>^2)$, and the heuristic scaling law $N=O(t^{-2/5})$ \cite{glasner2003coarsening}.
\par

\underline{\emph{Stage II: Drop-wise-condensing dynamics.}}
For $t\gg O(\beta^{-1})$, the remaining droplets are significantly larger in size with large peak-to-peak spacing $\lambda$ due to the early-stage coarsening (see Fig.~\ref{fig:D_lamb_dist}(left)). We have shown in Fig.~\ref{fig:twoDropTrajectories} that pairwise growth dynamics (region A) are the predominant events in this case, leading to more droplet collisions. Unlike the mass-conserving coarsening which preserves the total mass of droplets, the non-conservative coarsening modified by condensation is constrained by the finite domain size. When an array of large droplets with average half-width $\left<w(\pps)\right>$ are placed in the domain of size $L$, the number of droplets $N$ satisfies $2N\bra w(\pps)\ket  + N\bra D_A\ket =L$. Therefore, $N$ satisfies the constraint $N<L/(2\left<w\right>)$.
Using the estimate \eqref{eq:drop_width}, the droplet half-width $w\sim A/\pps$, and we arrive at $N = O (\left<1/\pps\right>^{-1})$. Applying Jensen's inequality, we get $\left<1/\pps\right>^{-1} \le  \left<\pps\right>$. Therefore,  we have an estimate for the relation between the number of droplets and their average pressure $N\propto \left<\pps\right>$.
This indicates that an estimate for the number of droplets in time can be obtained by analyzing the average droplet pressure in time.

\begin{figure}
    \centering
    \includegraphics[height=5cm]{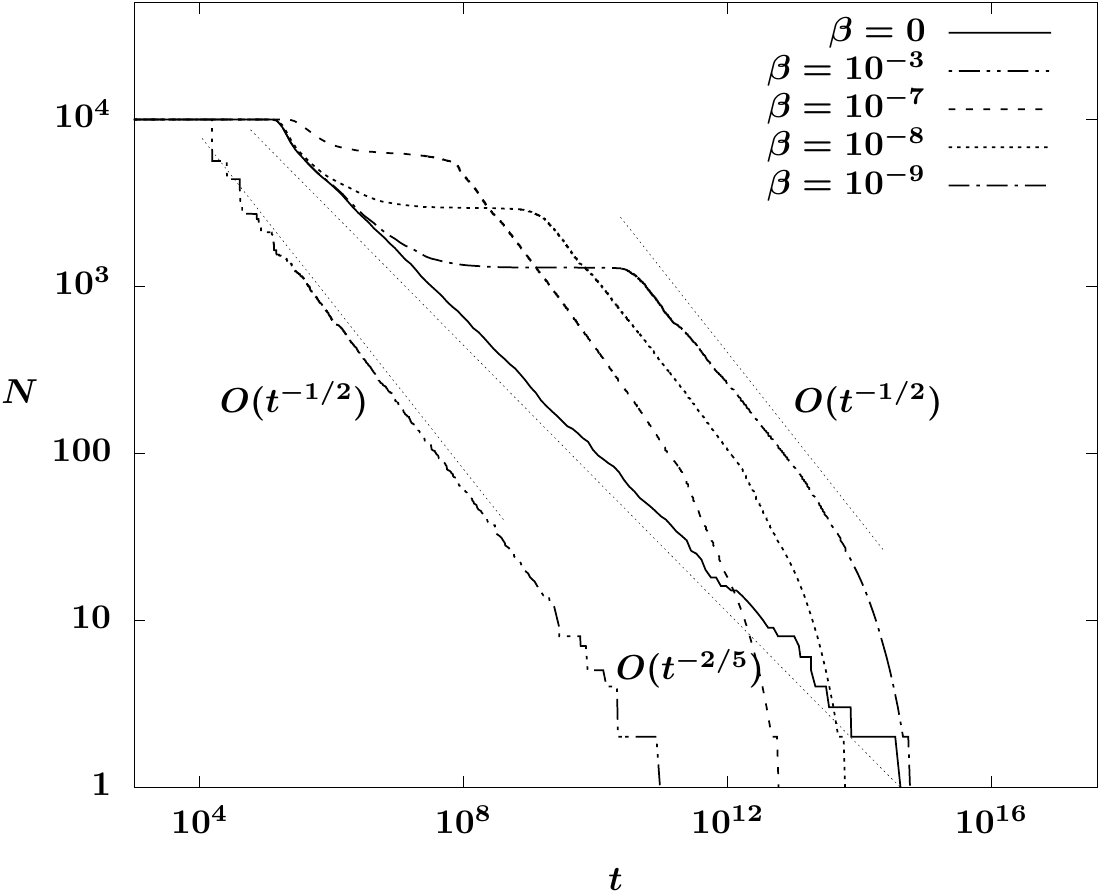}\quad
    \includegraphics[height=5cm]{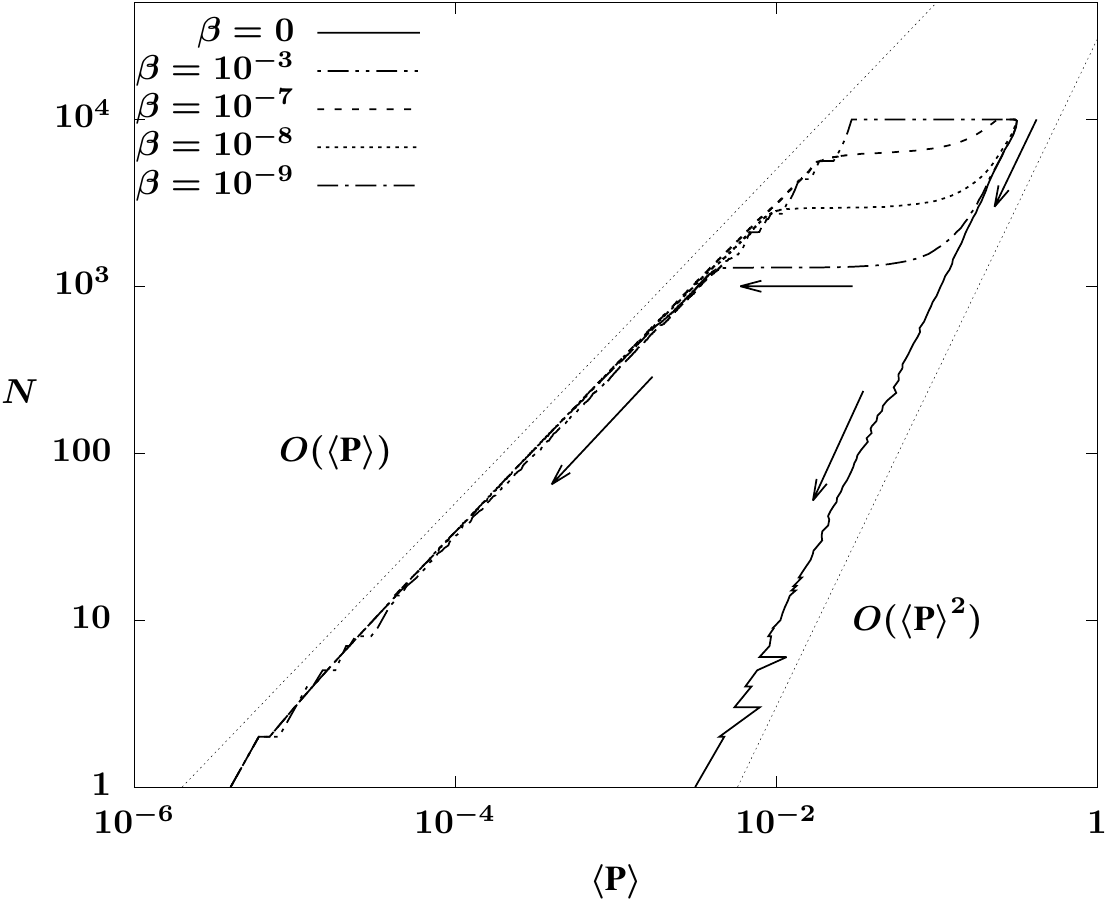}~
    \caption{(Left) Coarsening modified by weak condensation with $\beta=10^{-3}, 10^{-7}, 10^{-8}, 10^{-9}$ compared against the mass-conserving case with $\beta=0$, all starting from an identical initial droplet configuration with $N=10^4$ droplets. 
    (Right) Unlike the mass-conserving case with $\beta=0$ that follows the scaling $N = O( \left<\pps\right>^2)$, for $\beta = 10^{-7}, 10^{-8}, 10^{-9}$ the average droplet pressure $\left<\pps\right>$ shows a scaling transition from $N = O( \left<\pps\right>^2)$ to $N = O( \left<\pps\right>)$ in time as the number of droplets $N$ decreases (indicated by arrows). For $\beta=10^{-3}$, the full forms of the fluxes in the first expressions of \eqref{hmflux}--\eqref{hmflux2} are used and the coarsening dynamics follows the scaling $N = O(t^{-1/2})$.}
    \label{fig:N_P_relation}
\end{figure}

A sequence of numerical studies on coarsening dynamics confirms the regime transition from the early stage scaling law $N= O(\left<\pps\right>^2)$ to the later stage law $N=O(\left<\pps\right>)$.
Figure~\ref{fig:N_P_relation} compares the mass-conserving coarsening dynamics ($\beta=0$) with the coarsening dynamics modified by weak condensation effects ($\beta=10^{-9}, 10^{-8},10^{-7},10^{-3}$), all starting from an identical initial droplet configuration with $N=10^4$ droplets.  With the presence of weak condensation effects, the augmented dynamical system exhibits
a transition from the collapse-dominated $N=O( \bra\pps\ket^2)$ scaling law to the collision-dominated $N=O(\bra\pps\ket)$ coarsening over time, while the mass-conserving case $(\beta=0)$ remains close to the $N = O(\bra\pps\ket^2)$ scaling law  (see Fig.~\ref{fig:N_P_relation}~(right)).
During the transitional phase between the two stages, we observe that the number of droplets $N$ remains nearly constant for an extended period of time ($10^7\lesssim t \lesssim 10^{10}$ in the case of $\beta = 10^{-9}$), while the average pressure $\bra\pps\ket$ experiences a significant decrease (see the flat portion of the curves in Fig.~\ref{fig:N_P_relation}(right)). This observation suggests that the majority of droplets condense quasi-independently during this transitional regime, which is consistent with the $\mathcal{M} = O(t)$ scaling depicted in Fig.~\ref{fig:N_M_t}~(right). 
For the case with relatively strong condensation effects (see Fig.~\ref{fig:N_P_relation} for $\beta=10^{-3}$), we used the full forms of the fluxes in terms of hyperbolic functions, as given in the first expressions in \eqref{hmflux}--\eqref{hmflux2}. After early-stage ``stair-case" behaviors \cite{gratton2009transient}, the coarsening dynamics quickly transition into the regime following the scaling $N=O(t^{-1/2})$ and $N=O(\bra\pps\ket)$.
\par
To heuristically obtain a scaling law for the long-time behavior of an array of $N$ weakly condensing droplets in the spirit of mean-field models, we consider the dynamics of a single typical droplet with index $k$ at pressure $\pps_k$ in the array \cite{gratton2009transient,dai2010mean}. We assume that this droplet has both of its neighboring droplets at the average pressure, $\pps_{k\pm 1} =\pavg$, but retain the contact line distances $D_{k+1,k}$ and $D_{k, k-1}$ satisfying $(D_{k+1,k}^{-1}+D_{k,k-1}^{-1})/2 = \bra D_H\ket^{-1}$ and $(D_{k+1,k}+D_{k,k-1})/2 = \bra D_A\ket$. 
Here  $\bra D_H\ket$ and $\bra D_A\ket$ represent the local harmonic and arithmetic averages of the contact line distances.
Then from equation \eqref{eq:ODE_system}, we arrive at the evolution equation for $\pps_k$,
\begin{equation}
    {d\pps_k\over dt} = C_p(\pps_k)\left( {2\Hm^3\over \bra D_H\ket} (\pps_k -\pavg) +
    {\beta\mu\over 6}\bra D_A\ket ( 2\pavg +4 \pps_k -6\pstar) + \beta \Jnc(\pps_k)\right).
\label{eq:meanfield}
\end{equation}
We further assume that in the long-time condensation limit, $\pps_k \ll \pstar$ and $\pavg \ll \pstar$, and equation \eqref{eq:meanfield} reduces to
\begin{equation}
    {d\pps_k\over dt} = C_p(\pps_k)\left( {2\Hm^3\over \bra D_H\ket} (\pps_k -\pavg) -{\beta\mu}\bra D_A\ket  \pstar + \beta \Jnc(\pps_k)\right).
\lbl{eq:meanfield_reduced_0}
\end{equation}
Our numerical evidence in Fig.~\ref{fig:D_lamb_dist}~(right) suggests that in the later stage, the global harmonic average satisfies $\bra D_H\ket = O(1)$ and the arithmetic average satisfies $\bra D_A\ket \gg \bra D_H\ket$. Therefore, for large $t$, the leading-order equation of \eqref{eq:meanfield_reduced_0} becomes
\begin{equation}
    {d\pps_k\over dt} \sim \beta C_p(\pps_k)\left[-\mu\bra D_A\ket\pstar +  \Jnc(\pps_k)\right].
\lbl{eq:meanfield_reduced}
\end{equation}
Using the forms of $C_p(\pps_k)$ and $\Jnc(\pps_k)$ defined in \eqref{eq:coeff} and \eqref{eq:noncons_flux},
we take the condensation limit $\pps_k \to 0$ in \eqref{eq:meanfield_reduced} and arrive at the leading-order evolution equation for the droplet pressure
\begin{equation}
    \frac{d\pps_k}{dt} \sim -\frac{3\beta \pstar\pps_k^3}{4A^3} \left[\mu\bra D_A\ket-\frac{2}{A}\ln\left(\frac{K}{2A^2}\pps_k\right)\right].
\lbl{eq:coarsening_dpkdt}
\end{equation}
Also, since the right-hand-side of \eqref{eq:coarsening_dpkdt} is a concave
function of $\pps_k$, averaging over all the droplets in the system and applying Jensen's inequality to \eqref{eq:coarsening_dpkdt} yields
\begin{equation}
    \frac{d\left<\pps\right>}{dt} \lesssim -\frac{3\beta \pstar\left<\pps\right>^3}{4A^3} \left[\mu\bra D_A\ket-\frac{2}{A}\ln\left(\frac{K}{2A^2}\left<\pps\right>\right)\right].
\lbl{eq:coarsening_dpdt}
\end{equation}
\par
The two terms on the right-hand-side of \eqref{eq:coarsening_dpdt} correspond to the non-conservative fluxes arising from the inter-drop films and the core droplet condensation, both contributing to the overall mass increase.
The relative significance of the two fluxes further divides the later-stage dynamics into two distinct scenarios.
When the average contact line distance and the average pressure satisfy the relation $\mu\bra D_A\ket\gg \frac{2}{A}\ln\left(\frac{K}{2A^2}\left<\pps\right>\right)$, from \eqref{eq:coarsening_dpdt} we have
\begin{equation}
    \frac{d\left<\pps\right>}{dt} \propto -\left<\pps\right>^3 \qquad \implies \qquad
  N \propto  \left<\pps\right> = O(t^{-1/2}). 
\end{equation}
This transient behavior is observed in both Fig.~\ref{fig:N_M_t}(left) and Fig.~\ref{fig:N_P_relation}(left).
\par
As the coarsening and condensation dynamics continue, the droplet system is left with only a few extremely large droplets that occupy almost the entire domain.
In this case, we have $\mu\bra D_A\ket\ll \frac{2}{A}\ln\left(\frac{K}{2A^2}\left<\pps\right>\right)$, and the estimate
\eqref{eq:coarsening_dpdt} yields the leading-order governing equation for the average pressure
\begin{equation}
    \frac{d\left<\pps\right>}{dt} \propto \left<\pps\right>^3 \ln(\left<\pps\right>).
\end{equation}
Then based on the relation between the average pressure and the number of droplets, $N \sim L\left<\pps\right>/(2A)$, this estimate leads to a modified coarsening rate that satisfies
\begin{equation}
\frac{dN}{dt}\propto N^3\ln(cN),
\label{eq:N_log}
\end{equation}
where the constant $c\sim 2A/L$. Such behavior is also observed in Fig.~\ref{fig:N_M_t}(left) and in Fig.~\ref{fig:N_P_relation}(left),
where the number of droplets $N$ decays at a faster rate following the logarithmic law as $N \searrow 1$.
\par
Using the relation between the total mass of droplets and the average pressure $\mathcal{M} = N\bra M \ket\propto N\bra 1/\pps^2\ket \gtrsim N/\bra\pps\ket^2$, we also obtain an estimate for the total mass of droplets in the system.
That is, for the case $\mu\bra D_A\ket\gg \frac{2}{A}\ln\left(\frac{K}{2A^2}\left<\pps\right>\right)$, the total mass of droplets grows following $\mathcal{M} = O(\sqrt{t})$. For the final stage as $\bra\pps\ket \searrow 0$, with $\mu\bra D_A\ket\ll\frac{2}{A}\ln\left(\frac{K}{2A^2}\left<\pps\right>\right)$, 
the mass increases following a logarithmic law,
\begin{equation}
    \frac{d\mathcal{M}}{dt}\propto - \frac{1}{\mathcal{M}}\ln \left(\frac{\tilde{c}}{\mathcal{M}}\right),
\label{eq:M_log}
\end{equation}
where $\tilde{c}$ is a constant.
This is consistent with the observation shown in Fig.~\ref{fig:N_M_t}~(right).
\par 
\underline{\emph{Stage III: filmwise condensation dynamics.}} After the final droplet fills up the entire domain, the system undergoes filmwise condensation following the scaling law $\M(t)\sim L\sqrt{2\beta\pstar t}$ in \eqref{flood} 
(see the dashed curve in Fig.~\ref{fig:N_M_t}~(right)). This is similar to the filmwise condensation observed in Fig.~\ref{fig:SingleDropCondense} for $t>t_f$.

\section{Conclusions}
We have studied the coarsening dynamics of volatile thin liquid films from the PDE model \eqref{Mainpde} in the weak condensation limit $\beta\to 0$. A lower-dimensional dynamical system \eqref{eq:ODE_system} was derived for the behavior of interacting droplets parameterized by their positions and pressures. Using this reduced-order system, we have investigated various droplet dynamics and heuristically obtained scaling laws for different stages of long-time coarsening dynamics with weak condensation. Unlike the mass-conserving coarsening dynamics (as described by \eqref{Mainpde} with $\beta=0$) where the scaling law is primarily determined by droplet collapse events, the coarsening dynamics of weakly-condensing droplets exhibits a transition from a collapse-dominated stage to a collision-dominated stage.
\par
Heuristic scaling laws were presented for the coarsening dynamics with weak condensation. Rigorous justification of the arguments sketched out here are needed. For example, more thorough analysis should be carried out for the transition from the stage (I) to stage (II) in the coarsening dynamics.
For future work, we are interested in exploring the regimes of coarsening of thin films with stronger condensation effects (for $\beta=O(\eps^3)$ and larger).
Then more careful examination of phase change at droplet contact lines is needed, the full forms of the fluxes \eqref{hmflux} and \eqref{hmflux2} should be retained, 
and the thin films between drops have pressure profiles $p(x)$ with boundary layers, which would give rise to different forms for the dynamical system for droplet interactions.
Additionally, it would also be interesting to investigate how the coarsening dynamics are affected by the interplay between condensation, Marangoni effects, and gravity \cite{gratton2008coarsening}. We note that thin films with weak condensation may the simplest dynamic regimes for volatile films -- evaporation-dominated dynamics may require different analytical descriptions.
\par
Moreover, the mass-conserving droplet coarsening mechanism on a two-dimensional spatial domain has been explored in previous works \cite{glasner2009ostwald,pismen2004mobility}. We expect that extending our findings from this study to a two-dimensional setting will yield more intricate coarsening dynamics, offering a more accurate representation of the complex phenomena observed in real-world physical applications \cite{rose2002proc,enright2014dropwise,wen2017hydrophobic,wen2017wetting,anderson2012}.

\section{Acknowledgments}

TW acknowledges support from grant NSF DMS 2008255. HJ acknowledges support from grant NSF DMS 2309774 and the NCSU FRPD Program. We also appreciated very helpful comments and questions from the reviewers that enhanced and clarified the manuscript through revisions.

\bibliographystyle{abbrv}

\bibliography{bibliographHJi.bib}
\vfill

\end{document}